\documentclass{aa}
\usepackage[varg]{txfonts}
\usepackage{natbib}
\bibpunct{(}{)}{;}{a}{}{,}
\usepackage{graphicx}
\usepackage{caption}
\usepackage{subcaption}
\graphicspath{{./}}
\usepackage{float}
\floatstyle{plaintop}
\restylefloat{table}
\usepackage{xcolor}
\usepackage[colorlinks=true,     linkcolor=blue, citecolor=blue, filecolor=blue, urlcolor=blue]{hyperref}

\begin{document}

    \title{Parsec scales of carbon chain and complex organic molecules in AFGL 2591 and IRAS 20126}

   \author{P. Freeman
          \inst{1}
          \and
          S. Bottinelli 
          \inst{2}
          \and
          R. Plume
          \inst{1}
          \and
          E. Caux
          \inst{2}
          \and
          C. Monaghan
          \inst{1}
          \and
          B. Mookerjea
          \inst{3}
          }

   \institute{University of Calgary, Calgary, Canada\\ \email{pamela.freeman@ucalgary.ca}\\
         \and
             IRAP, Université de Toulouse, CNRS, UPS, CNES, Toulouse, France\\
        \and
            Tata Institute of Fundamental Research, Mumbai, India
             }

\date{Submitted 22 June 2023; Revised 04 August 2023; Accepted 08 August 2023}

\abstract{There is a diverse chemical inventory in protostellar regions leading to the classification of extreme types of systems. Warm carbon chain chemistry sources, for one, are the warm and dense regions near a protostar containing unsaturated carbon chain molecules. Since the presentation of this definition in 2008, there is a growing field to detect and characterise these sources. The details are lesser known in relation to hot cores and in high-mass star-forming regions -- regions of great importance in galactic evolution.}{To investigate the prevalence of carbon chain species and their environment in high-mass star-forming regions, we have conducted targeted spectral surveys of two sources in the direction of Cygnus X -- AFGL 2591 and IRAS 20126+4104.}
{We observed these sources in frequency ranges around 85, 96, and 290 GHz with the Green Bank Telescope and the IRAM 30m Telescope. We have constructed a Local Thermodynamic Equilibrium (LTE) model using the observed molecular spectra to determine the physical environment in which these molecules originate. We map both the observed spatial distribution and the physical parameters found from the LTE model. We also determine the formation routes of these molecules in each source using the three-phase NAUTILUS chemical evolution code.}
{We detect several lines of propyne, CH$_3$CCH, and cyclopropenylidene, $c$-C$_3$H$_2$ as tracers of carbon chain chemistry, as well as several lines of formaldehyde, H$_2$CO, and methanol, CH$_3$OH, as a precursor and a tracer of complex organic molecule chemistry, respectively. We find excitation temperatures of 20-30~K for the carbon chains and 8-85~K for the complex organics. The observed abundances, used as input for the chemical evolution code, are 10$^{-9}$ to 10$^{-10}$ for both CH$_3$CCH and CH$_3$OH. The CH$_3$CCH abundances are reproduced by a warm-up model, consistent with warm carbon chain chemistry, while the observed CH$_3$OH abundances require a shock mechanism sputtering the molecules into the gas phase.}{Single-dish observations are useful for studying the envelope-scale chemistry of star-forming regions, including mechanisms such as warm carbon chain chemistry. As well, LTE models lend well to the wide-band maps obtained from these telescopes. The physical and chemical environment determined for complex hydrocarbons and complex organics lends understanding to high-mass star formation.}

\keywords{stars: formation
 -- ISM: molecules
 -- Astrochemistry -- Submillimeter: ISM
 }

\maketitle

\section{Introduction}\label{section:intro}

Molecular clouds are the sites of star formation, where gas collapses into dense cores before forming protostars. The chemical makeup of molecular clouds therefore leads to that of protostellar systems. Observing the chemical complexity of star formation and modelling its evolution provides an invaluable link between the interstellar medium and planetary composition. In the last few decades, the diverse chemical makeup of star-forming regions has been revealed (\citealt{herbst2009, sakai2013, jorgensen2020}). 

There are two major carbon-based groups commonly studied in these regions: interstellar complex organic molecules (iCOMs) and carbon chain molecules (CCMs). Complex, in interstellar terms, means any molecule with six or more atoms. COMs and CCMs are important to observe as their formation is sensitive to environmental conditions, thus can supplement existing information about interstellar objects traced by simpler, often studied, molecules such as CO, HCO+, or HCN.

COMs, which are carbon-bearing saturated complex molecules, have been observed in abundance in the hot (T $>$ 100 K) and dense ($n_{\text{H}_2}$ $>$ 10$^6$ cm$^{-3}$) cores of star-forming regions (see \citealt{herbst2009} for a review of these molecules). The hot cores of high-mass star formation are well known (high mass stars have M $>$ 8 M$_\odot$; \citealt{blake1987, caselli1993, macdonald1996, helmich1997}) as are their compact ($\le$ 100 AU in radius) equivalent in low-mass regions, dubbed hot corinos (\citealt{Cazaux2003, Ceccarelli2004, Bottinelli2004, Bottinelli2007}). In star-forming regions complex molecules are formed on and released from interstellar grain surfaces. This chemistry is initiated by hydrogenation of CO, depleted onto grain surfaces from the gas phase, to make CH$_3$OH. COM formation in the gas phase is dominated by ion-neutral reactions, though the formation of saturated molecules in cold phases is inefficient \citep{herbst2009}. 

CCMs, on the other hand, are unsaturated hydrocarbons (e.g. C$_n$H, HC$_n$N) that are well-known in cold (T $\le$ 10 K) molecular clouds and starless cores such as Sagitarrius B2, Taurus Molecular Cloud, and Lupus-1A (\citealt{avery1976, broten1978, kroto1978, little1978, sakai2010}). The cold starless core phase forms CCMs in the gas phase efficiently prior to the uptake of C in CO, a stable molecule. CCMs, then, were thought to have a short lifetime -- present in the early stages of cloud core evolution but becoming deficient in later stage star-forming regions.

However, another formation route for CCMs is `Warm Carbon Chain Chemistry' (WCCC) (\citealt{sakai2008, aikawa2008, sakai2013}). If atomic carbon depletes onto interstellar grains early in the star formation process, successive hydrogenation of C on the grain surface forms CH$_4$. As the gas warms in the star formation process and CH$_4$ is liberated from dust grains, it acts as the precursor for CCM formation. The condition keeping carbon in its atomic form prior to depleting onto grain surfaces is thought to be a short timescale for collapse \citep{sakai2008} or UV radiation (\citealt{spezzano2016, higuchi2018}). A WCCC source is characterised by two conditions: that there is an abundance of various carbon-chain molecules, and these molecules are concentrated in the warm and dense part around a protostar \citep{sakai2013}. Following the detection of the low-mass star-forming region L1527 as a WCCC source \citep{sakai2008}, WCCC characteristics have also been seen in a high-mass star-forming region, a giant H\,{\footnotesize II} region, and a starless core (\citealt{mookerjea2012, saul2015, wu2019}).

Hot corinos and warm carbon chain chemistry (WCCC) sources now classify two extreme types of protostellar systems, due to this chemical differentiation. However, many sources have characteristics of both types -- a hybrid source. The second WCCC source reported by \citet{sakai2009}, IRAS 15398$-$3359 in Lupus, hosts a hot corino \citep{okoda2023}. iCOMs are emitted on a scale 10-100$\times$ smaller than the CCM emission. \citet{okoda2023} suggest the hybrid nature supports both the grain mantle composition and the physical environment influencing the chemistry. L483, a Class 0 protostar, shows WCCC characteristics on a 1000 AU scale and also hot corino characteristics on ($<$) 100 AU scales (\citealt{sakai2009,oya2017}). The Bok globules B335 and CB 68, which are useful environments to study isolated protostellar sources, show hot corino emission on tens of AU scales and also carbon chain emission on hundreds to a thousand AU scale (\citealt{imai2016,Imai2022}). \citet{higuchi2018} find numerous intermediate sources in the Perseus Molecular Cloud, though find high CCH/CH$_3$OH ratios -- characteristics of WCCC -- towards isolated or edge-of-the-cloud sources owing to both differences in timescales and effects of UV radiation.

Carbon chain chemistry, traced by C$_4$H (\citealt{Lindberg2016, graniger2016}), CCH (\citealt{higuchi2018, bouvier2020}), and cyanopolyynes (\citealt{taniguchi2018b, taniguchi2021}), may be compared to complex organic chemistry, usually traced by CH$_3$OH, through parameters such as line density, column density, spatial distribution, or temperature. These studies identify that CCMs or their tracers are often in cooler, more extended gas, such as the envelope around hot cores, produced from CH$_4$ after it is sublimated from dust grain surfaces. There are dissimilarities between CCMs and COMs in spatial distribution, line profiles, and column densities. \citet{Lindberg2016}, referencing \citet{rodgers2001}, are careful to note that certain classes of CCMs may have different correlations to methanol.

In this paper, we examine tracers of warm carbon chain and complex organic chemistry in the high-mass star-forming regions AFGL 2591 and IRAS 20126. Section~\ref{section:study} describes the molecules and sources. Section~\ref{section:observations} describes the observations and Sect.~\ref{section:results} the detected lines and Local Thermodynamic Equilibrium (LTE) model results. Section~\ref{section:discussion} discusses the results in wider context, and Sect.~\ref{section:summary} provides a summary.

\section{Objects of study}
\label{section:study}

\subsection{Warm carbon chain molecules -- propyne and cyclopropenylidene}

Propyne, or methyl acetylene, CH$_3$CCH, is part of the methylpolyyne family \citep{irvine1981}. It is a symmetric rotor with many transitions closely spaced in frequency. Numerous lines are thus captured in a small bandwidth, and are useful for analysis of the physical conditions, especially temperature, in LTE (\citealt{Askne1984, kuiper1984, Bergin1994}). It was first detected in the giant molecular clouds Sgr B2 and Orion A, two sites with active star formation (\citealt{snyder1973, lovas1976}). It continues to be detected in low- and high-mass star-forming regions and is used as a tracer of chemical complexity (\citealt{Cazaux2003, taniguchi2018b, gianetti2017, santos2022}).

CH$_3$CCH can form both on grain surfaces and in the gas phase, with the former route directing observed abundances \citep{calcutt2019}. On grain surfaces, CH$_3$CCH develops from the successive hydrogenation of C$_3$ (\citealt{hickson2016, ines2018}). In the gas phase, the primary route is through ion-neutral reactions (\citealt{schiffbohme1979, taniguchi2019}) with contributions from neutral-neutral reactions \citep{turner1999}. Recently, studies note the importance of CH$_4$, forming on then desorbing from grain surfaces in the warm-up phase, and producing larger hydrocarbons through gas-phase ion-neutral reactions (\citealt{sakai2013, calcutt2019, taniguchi2019}). These hydrocarbons dissociatively recombine to form CH$_3$CCH.

\citet{thaddeus1985} identified cyclopropenylidene, $c$-C$_3$H$_2$, as the first astronomically observed organic ring molecule. The follow up in \citet{Vrtilek1987}, reports on detections in Sgr B2, Orion, and TMC-1. $c$-C$_3$H$_2$ is formed in dense and `chemically young' gas, where carbon is in the atomic form \citep{spezzano2016}. Gas-phase ion molecule, H-atom transfer, and electron recombination reactions lead to its production (\citealt{park2006, sakai2013}). As this process starts with CH$_4$, desorbed from dust grains at 25~K, this molecule can be representative of WCCC. \citet{Aikawa2020} showed that with a new multi-layered ice mantle model abundances of c-C$_3$H$_2$ correlate with CH$_4$. 

\subsection{Complex organic molecules -- formaldehyde and methanol}

Formaldehyde, H$_2$CO, is not technically a complex species; however, it is chemically associated with larger organic molecules and we include it as a COM precursor in this paper. H$_2$CO was the first polyatomic organic molecule detected in space, \citet{Snyder1969} observed absorption spectra towards both galactic and extragalactic sources. Methanol, CH$_3$OH, is the simplest alcohol molecule. It has many low energy levels due to the torsional motion it undergoes as an asymmetric top molecule \citep{ball1970}. Methanol is an abundant complex molecule with numerous transitions in the mm and sub-mm range which are useful, and commonly used, as a probe of the physical conditions of an interstellar region. 

H$_2$CO and CH$_3$OH are both formed by the successive hydrogenation of CO after it freezes out onto the surface of grains in the cold cloud phase (\citealt{Charnley1997, garrod2013}). The hydrogenation of CO and H$_2$CO require activation energy, suggesting not all molecules of these species will be converted into larger molecules. Still, this is a competitive process even at low temperatures; these species are discovered in a variety of sources such as cold clouds, hot cores, outflows, shocks, the Galactic centre, and external galaxies (see \citealt{herbst2009} and references therein). These species are observed in both gas and solid phases and are key precursor molecules to larger COMs. 

Certain CH$_3$OH transitions are masers, pumped into an excited state either collisionally (Class I) or by far-infrared radiation (Class II). CH$_3$OH masers are detected in IRAS 20126 (Class I and II) and AFGL 2591 (Class II) \citep{batrla1988, Plambeck1990, harvey2008, moscadelli2011, Rodriguez2017, rygl2012}.

\subsection{Sources}

\begin{figure}[h!]
\centering
\includegraphics[width=0.5\textwidth]{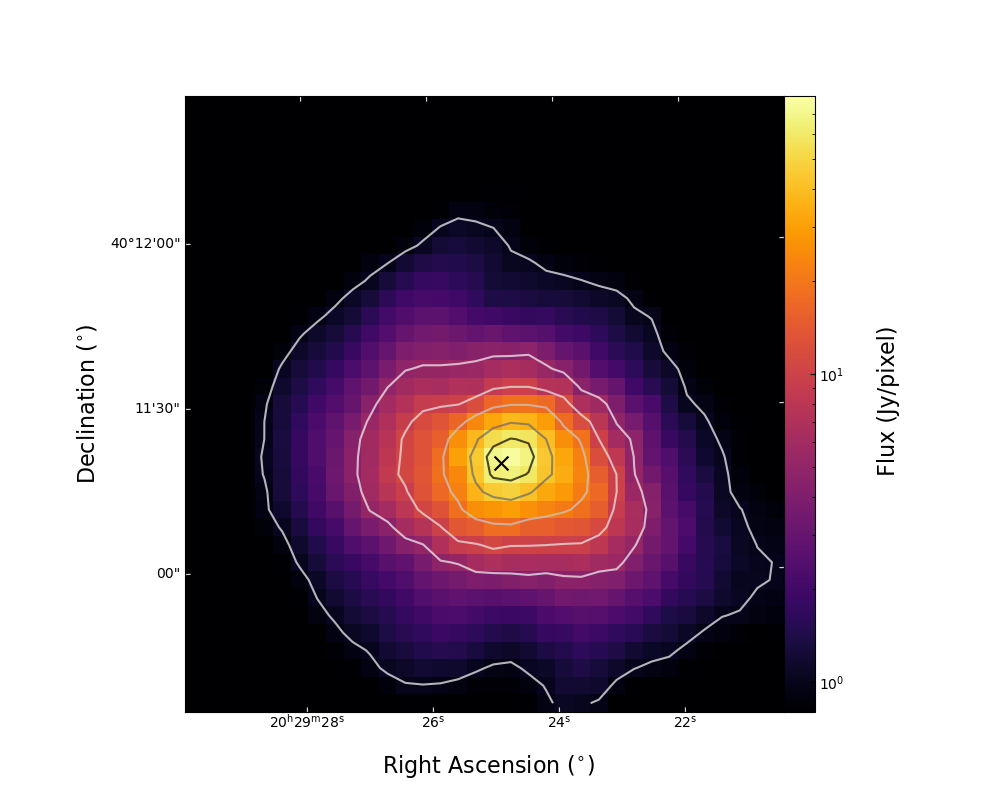}
\caption{Herschel PACS 160 $\mu m$ image of AFGL 2591 (proposal ID KPGT\_fmotte\_1, PI F. Motte). The colour scale is logarithmic from 0.8 to 80 Jy/pixel, while the contours are 1, 5, 10, 30, 40, and 60 Jy/pixel. The black `x' represents the dominant protostellar source VLA 3.}
\label{fig:afgl2591pacs}
\end{figure}

AFGL 2591 is the prototypical object in which to study physical and chemical processes during high-mass star formation. Figure~\ref{fig:afgl2591pacs} displays a Herschel PACS 160~$\mu m$ continuum image of the source. At a distance of 3.33$\pm$0.11 kpc \citep{rygl2012}, AFGL 2591 has a total mass of about 2$\times10^4$ M$_\odot$ and a measured total IR luminosity of 2$\times10^5$ L$_\odot$ \citep{sanna2012}. Within this clump there are several radio continuum sources. VLA 1 and 2 are optically thin H\,{\footnotesize II} regions \citep{trinidad2003}. VLA 3 is believed to be the youngest (dynamical age of about 2$\times10^4$ yr; Doty et al. 2002; Stäuber et al. 2005) and most massive source in the cluster, with an estimated mass of 38~M$_\odot$ \citep{sanna2012}. VLA 4 and 5 are faint sources detected in cm continuum observations \citep{johnston2013}. A large-scale east-west bipolar outflow powered by VLA 3 is seen in molecular observations which extends past the warm and dense envelope of the source (\citealt{lada1984, Hasegawa1995, sanna2012, johnston2013}). The simplest COMs and CCMs (CH$_3$OH, H$_2$CO, and CCH) are reported in AFGL 2591 via the James Clerk Maxwell Telescope (JCMT) Spectral Legacy and the CHESS surveys (\citealt{vanderwiel2011, kazmierczak2014}). However, these surveys are at high frequencies ($>$ 300 GHz) where COMs and CCMs do not necessarily emit their strongest lines. Thus, little is known about the relative abundance of COMs and CCMs in this source.

\begin{figure}[h!]
\centering
\includegraphics[width=0.5\textwidth]{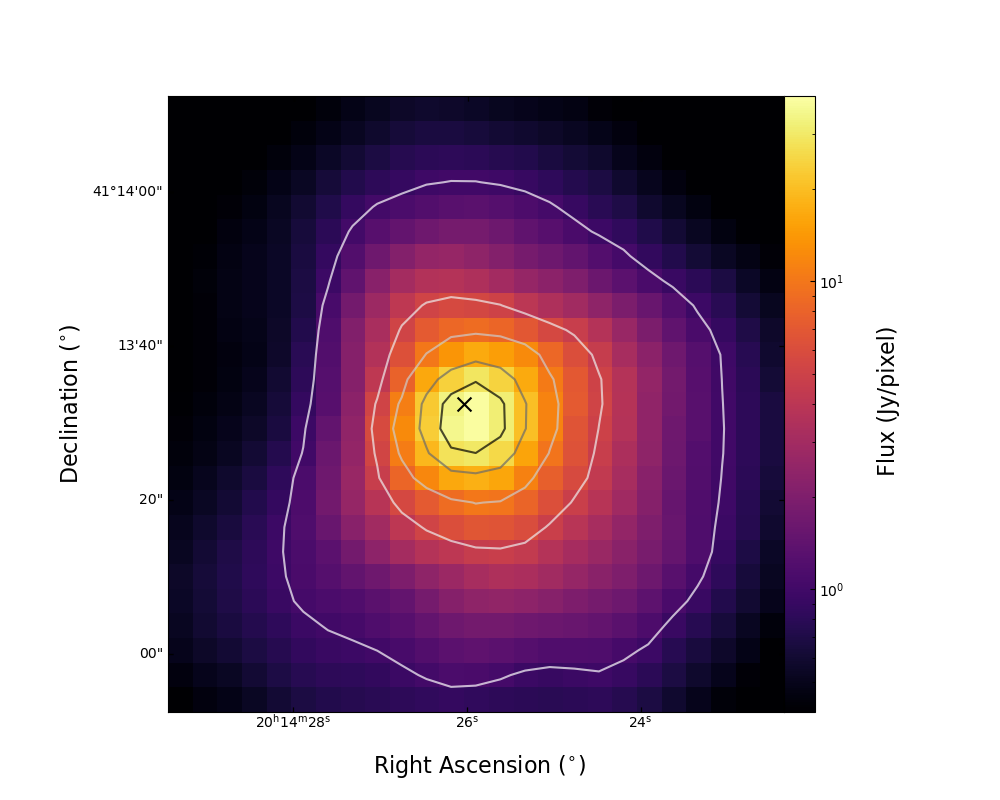}
\caption{Herschel PACS 160 $\mu m$ image of IRAS 20126+4104 (proposal ID OT1\_rcesaron\_1, PI R. Cesaroni). The colour scale is logarithmic from 0.4 to 40 Jy/pixel, while the contours are 1, 5, 10, 20, and 30 Jy/pixel. The black `x' represents the protostellar source.}
\label{fig:iras20126pacs}
\end{figure}

IRAS 20126+4104 (hereafter IRAS 20126) is a well-studied massive protostar that allows us to observe the early stages of massive star formation in a relatively simple system. Figure~\ref{fig:iras20126pacs} displays a Herschel PACS 160~$\mu m$ continuum image of the source. In the Cygnus X region at a distance of 1.6~kpc, IRAS 20126 has a luminosity of 1.3$\times10^4$ L$_\odot$ \citep{moscadelli2011}. The natal dense clump, which is a site of massive star formation, is isolated with a temperature of 40~K \citep{shepherd2000}. Within this clump observations of C$^{34}$S by \citet{cesaroni2005} confirm the presence of a large Keplerian disk with a radius of 5000~AU. Near-infrared K, L', and M' band observations by \citet{sridharan2005} with the United Kingdom Infrared Telescope, however, resolve the central source and suggest the presence of a more compact disk (R $\sim$ 1000 AU). This was confirmed by \citet{chen2016} whose modelling of SMA observations of CH$_3$CN confirm a 1.5~M$_\odot$ disk with a radius of 850~AU rotating about a 12~M$_\odot$ protostar. Numerous COMs have been detected in JCMT and Plateau de Bure Interferometer (PdBI) observations \citep{isokoski2013, palau2017}  which have also included the detection of some simple CCMs (CCH, CH$_3$CCH). However no surveys targeting CCMs have been reported. Thus, little is known about the relative abundance of COMs versus CCMs in this source as well.

\section{Observations}
\label{section:observations}

The IRAM 30-m observations were completed November 2020 and April 2021 (project codes 021-20 and 122-20, PIs S. Bottinelli and R. Plume). Using the on-the-fly observational mode with position switching, we mapped a $1'\times1.5'$ region in both AFGL 2591 and IRAS 20126 in the frequency range 287.22 - 295~GHz. The offset position was -120$\arcsec$ horizontally and -240$\arcsec$ vertically. The phase centre was $\alpha$J(2000)$=20^\mathrm{h}29^\mathrm{m}24\fs 9$ and $\delta$J(2000)$=+40\degr 11\arcmin20\farcs0$ for AFGL 2591 and $\alpha$J(2000)$=20^\mathrm{h}14^\mathrm{m}26\fs 04$ and 
$\delta$J(2000)$=+41\degr 13\arcmin32\farcs5$ 
for IRAS 20126. The EMIR receiver, in bands E1 and E3, was connected to the FTS200 backend (Fast Fourier Transform Spectrometer at 200 kHz resolution). The range 131.2 - 138.98~GHz was observed simultaneously in this set-up, but will not be discussed in this paper due to the much larger beam size. The pointing, checked every 90 minutes or less, was done using either K3-50A or NGC 7027 and resulted in typical corrections of 1-2$^{\prime\prime}$. The focus, completed every three hours or after a sunset or sunrise, was done with the same source as for pointing and resulted in typical corrections of $<$~0.3$\arcsec$. The atmospheric opacity, as measured by the 225~GHz taumeter, aside from part of the day November 28, 2020, was $<$~0.3, and often $<$~0.1. The system temperatures observing IRAS 20126 were 350-500~K, and observing AFGL 2591 were 300-350~K. The beam size is 9.3$\arcsec$ at 290~GHz, with a pixel size of 4.4$\arcsec$. Data reduction, and the production of maps, was completed with GILDAS/CLASS\footnote{https://www.iram.fr/IRAMFR/GILDAS}. 

The 100m Green Bank Telescope observations were completed in March 2021, in the frequency ranges 84.5-85.75~GHz and 95.55-96.8~GHz using all 16 beams of the ARGUS focal plane array and the VErsatile GBT Astronomical Spectrometer (VEGAS) spectral line backend (project code 21A-039, PI P. Freeman). We obtained 1$\arcmin$ DAISY on-the-fly maps for both frequency ranges in AFGL 2591 and IRAS 20126, centred as above, utilising position-switching to a location +3$\arcmin$ in azimuth for the reference measurements. Each petal, with a map radius of 0.5$\arcmin$, took 54.7 seconds with a scan rate of 2.3$\arcsec$ per second and an integration time of 1 second. The pointing and focus was completed every 30-40 minutes with the X-band receiver using the GBT's automatic scan AutoPeakFocus on source 2015+3710. At the ARGUS frequencies, the automatic scan AutoOOF was used to correct for errors in the reflecting surface using source 1229+0203. The system temperatures ranged from 100-200~K. With spectrometer mode 2 we obtained a spectral sampling of 92~kHz in both frequency ranges, or 0.32~km s$^{-1}$~at 85~GHz and 0.28~km s$^{-1}$~at 96~GHz. The beam size is 9.2$\arcsec$ or 10.0$\arcsec$ at 96 or 85~GHz respectively, with a pixel size of 2$\arcsec$. The data were reduced and calibrated using GBTIDL\footnote{https://gbtidl.nrao.edu/index.shtml}.

The spectral axis of the IRAM data was smoothed from the native spectrometer resolution to reduce the high levels of noise. The final spectral sampling is 781~kHz, or 0.80~km s$^{-1}$. Given the widths of the spectral lines, this is still adequate to confidently detect and fit these lines. The GBT 85 GHz data was spatially smoothed by a factor of two in pixel size to improve the noise levels and to closer match the IRAM pixel size. With this, and the higher quality GBT observations, the GBT data was not spectrally smoothed as the noise levels are lower. The GBT 96 GHz and IRAM 290 GHz data were regridded, smoothed, and rebinned in CASA\footnote{https://casa.nrao.edu/} to match the 85 GHz beam size and spatial grid. The final restored beam is 10.0$\arcsec$ with a pixel size of 4$\arcsec$. Average rms noise levels are shown in Table~\ref{tab:obs}. In the Results, Sect.~\ref{section:results}, the data are in units of T$_A$*, adjusted during calibration. In the LTE Model, the data are converted to T$_{\rm mb}$ using the telescope B$_{\rm eff}$/F$_{\rm eff}$ values provided by the respective observatories: GBT at 85~GHz, 0.4545; GBT at 96~GHz, 0.3838; IRAM EMIR at 290~GHz, 0.547.\\

\medskip

\begin{table*}
\caption{Details of the data set used in this paper.}
\label{tab:obs}
\centering
\begin{tabular}{ l c c c c c }
\hline\hline
Telescope & Frequency & Spectral Sampling &
Spatial Sampling &  IRAS 20126 RMS noise & AFGL 2591 RMS noise\\
 & [GHz] & [km s$^{-1}$] &
[$\arcsec$] & [T$_A$*, mK]  & [T$_A$*, mK] \\
\hline
 GBT & 84.50-85.75 & 0.32 & 4 & 27.20 & 25.91 \\
 GBT & 95.55-96.80 & 0.28 & 4 & 25.99 & 23.61 \\
  IRAM & 287.00-295.00 & 0.80 & 4 & 37.49 & 75.28 \\
\hline
\end{tabular}
\tablefoot{The spectral sampling is the channel width and the spatial sampling is the pixel size.}
\end{table*}

\section{Results}
\label{section:results}

\subsection{Detected lines}

Species were identified through the  CASSIS\footnote{http://cassis.irap.omp.eu/} LTE+RADEX\footnote{http://cassis.irap.omp.eu/help/?page=html/m\_lte\_radex} module. First, the CDMS and JPL catalogues under the `Species' tool in CASSIS allowed us to determine possible detected species across the observed spectrum. These two catalogues are cross-referenced for confirmation. Next, we used the LTE Model in the module to model the spectrum. We input an estimated column density, excitation temperature, FWHM line width, source size and $v_{LSR}$. The spectrum was computed across the frequency range using telescope parameters specific to IRAM and the GBT as documented within CASSIS. Line identification was then done visually, matching the model to the data. It is important to note that this model is only for line identification and not to accurately determine the physical conditions.

The combined GBT and IRAM data set contain several simple and complex molecules. The complete spectral surveys are to be described in Freeman et al. (in prep). Table \ref{tab:lines} lists all possible CH$_3$CCH (CDMS tag 40502 for the vibrational ground state), $c$-C$_3$H$_2$ (CDMS tag 38508), CH$_3$OH (CDMS tag 32504), and H$_2$CO (CDMS tag 30501) lines with A$_{\mathrm{ij}}$ $>$ 1$\times 10^{-6}$ s$^{-1}$ and E$_{\mathrm{up}}$ $<$ 150 K. For CH$_3$OH the A$_{\mathrm{ij}}$ limit was set to 2$\times10^{-6}$ s$^{-1}$ to remove the 40 K maser line. For CH$_3$CCH the E$_{\mathrm{up}}$ limit was increased to 200 K to account for the a-CH$_3$CCH line that has E$_{\mathrm{up}}$ $<$ 150 K, but has a higher tabulated value when the a- and e- types are combined in CDMS. This range was selected to restrict the catalogue transitions to the ones that could be detected with the sensitivity of our observations. The species types --  a- and e-CH$_3$CCH, o- and p-$c$-C$_3$H$_2$, A- and E-CH$_3$OH, o- and p-H$_2$CO -- were not differentiated as we do not detect enough lines of each. The VASTEL database information for each type is similar to that of CDMS, aside from including higher E$_{\mathrm{up}}$ a-CH$_3$CCH lines, and we expect the results to not be affected drastically by using the combination of the types. 

\begin{figure*}[h!]
\centering
\includegraphics[width=\textwidth]{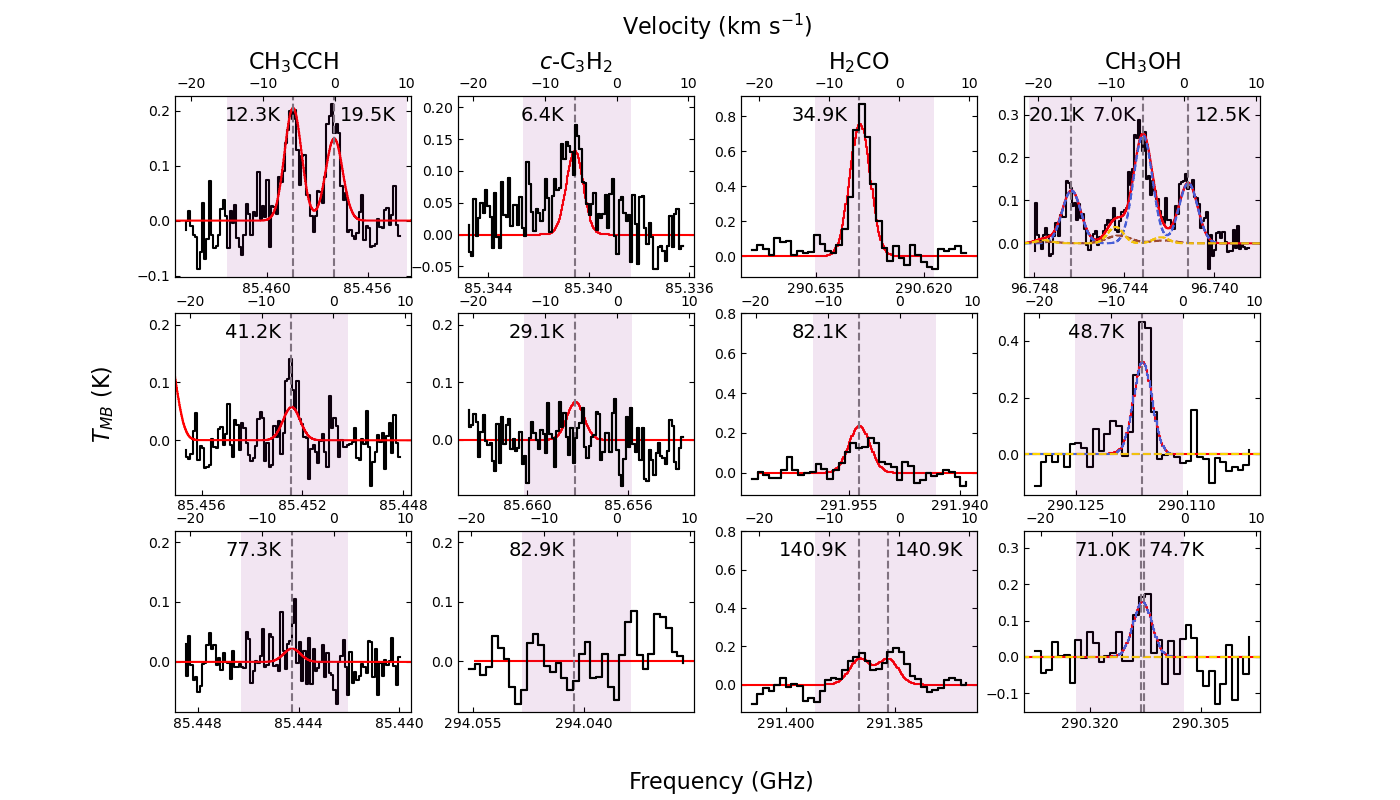}
\caption{Select transition lines of the carbon chain molecules, CH$_3$CCH and $c$-C$_3$H$_2$, the complex organic molecules CH$_3$OH, and the precursor complex organic molecule H$_2$CO, at the central pixel of the AFGL 2591 GBT and IRAM maps. The chosen lines represent a range of different upper energy levels, which are indicated in black on the plot. The dashed grey line represents the observed transition frequency at this location. The observed data is shown in black. The model spectra, to be discussed in Sect.~\ref{section:ltemodel}, are shown in colour. The red solid line represents the total model, and the sole component if only one component is fit. For the multi-component fit in CH$_3$OH the separate components are shown in blue, yellow, and brown dashed lines.}
\label{fig:afgl2591spectra}
\end{figure*}

\begin{figure*}[h!]
\centering
\includegraphics[width=\textwidth]{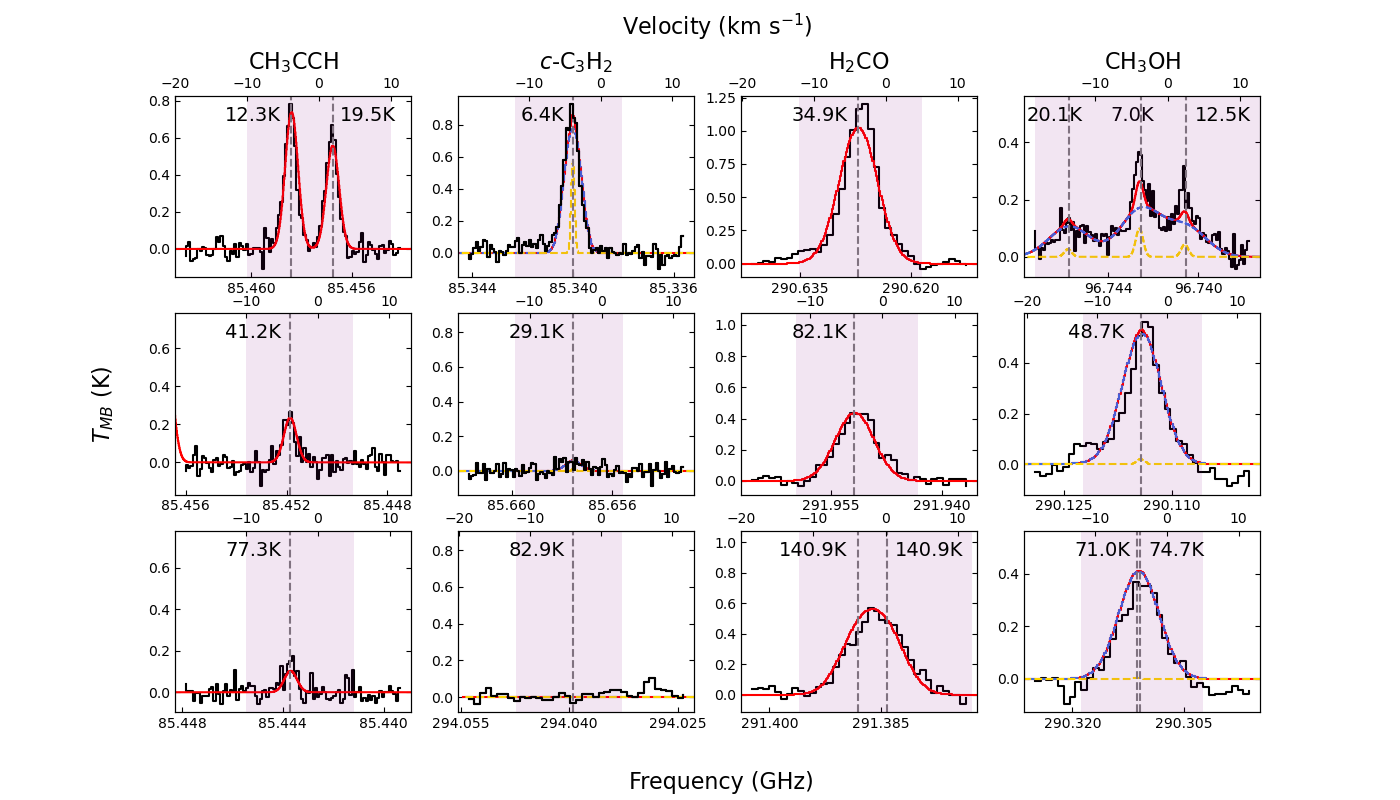}
\caption{Same as Fig.~\ref{fig:afgl2591spectra} for the central pixel of the IRAS 20126 GBT and IRAM maps. For the multi-component fits in $c$-C$_3$H$_2$ and CH$_3$OH the separate components are shown in blue and yellow dotted lines.}
\label{fig:iras20126spectra}
\end{figure*}

Figures~\ref{fig:afgl2591spectra} and \ref{fig:iras20126spectra} show examples of the spectra at the brightest pixel for the CH$_3$CCH 5$_0$-4$_0$ (E$_{\mathrm{up}}$\,=\,12 K) line and the CH$_3$OH 2$_0$-1$_0$, 2$_{-1}$-1$_{-1}$, 2$_0$-1$_0$ triplet (E$_{\mathrm{up}}$\,=\,7, 13, 20~K). Figures~\ref{fig:afgl2591ccmresults}, \ref{fig:afgl2591comresults}, \ref{fig:iras20126ccmresults}, and \ref{fig:iras20126comresults} display all transition lines of each molecule. The propyne detections with E$_{\mathrm{up}}$ from 10-30~K are easily detected and well-defined; rarely showing blended or complicated structure. The methanol lines from six to several tens of K are clearly detected. The higher energy lines are often blended -- in part due to methanol's complicated line structure -- and often show structure that is not represented by a single Gaussian. The IRAM data also suffer from higher noise due to unstable weather conditions and higher observed frequencies. These trends are consistent across both IRAS 20126 and AFGL 2591, with the latter displaying more complicated methanol line profiles.
\\

\medskip

\begin{table*}
\caption{Rotational transition lines of selected molecules in the observed frequency ranges.}
\label{tab:lines}
\centering
\begin{tabular}{ l l l c c c l l c c c }
\hline\hline
Molecule & \multicolumn{2}{c}{Quantum Numbers} & Frequency & $E_{\rm up}$ &  $A_{\rm ij}$ & \multicolumn{2}{c}{Quantum Numbers} & Frequency &
$E_{\rm up}$ & $A_{\rm ij}$ \\
 & \multicolumn{2}{c}{} & [GHz] & [K] & [10$^{-5}$ s$^{-1}$] & \multicolumn{2}{c}{} & [GHz] & [K] & [10$^{-5}$ s$^{-1}$]
\\
\hline
CH$_3$CCH & 5$_0$-$4_0$ & {e} & 85.4573 & 12.31 & 0.2
            & 17$_0$-17$_0$ & {e} & 290.5020 & 125.50 & 8.5\\
             & 5$_1$-4$_1$ & {e} & 85.4557 & 19.53 & 0.2
             & 17$_1$-17$_1$ & {e} & 290.4965 & 132.73 & 8.5\\
             & 5$_2$-4$_2$ & {e} & 85.4508 & 41.21 & 0.2
             & 17$_2$-17$_2$ & {e} & 290.4799 & 154.40 & 8.4\\
             & 5$_3$-4$_3$ & {a} & 85.4426 & 77.34 & 0.1
             & 17$_3$-17$_3$ & {a} & 290.4522 & 190.52 & 8.3 \\
            \hline
$c$-C$_3$H$_2$ & 2$_{1,2}$-1$_{0,1}$ & {o} & 85.3389 & 6.45 & 2.3
             & 4$_{3,2}$-4$_{2,3}$ & {o} & 85.6564 & 29.07 & 1.5 \\
            & 3$_{2,2}$-3$_{1,3}$ & {p}  & 84.7277 & 16.14 & 1.0 
            & 7$_{5,3}$-6$_{6,0}$ & {p} & 294.0356 & 82.91 & 2.4\\
            \hline
CH$_3$OH  & 2$_{0,2,0}$-1$_{0,1,0}$ & {A+} & 96.7413 & 6.96 & 0.3 
          & 4$_{3,2,1}$-5$_{2,4,1}$ & {E} & 288.7056 & 70.93 & 1.8\\
          & 2$_{1,2,2}$-1$_{1,1,2}$ & {E} & 96.7394 & 12.54 & 0.3 & 6$_{2,5,1}$-5$_{2,4,1}$ & {E} & 290.3077 & 71.01 & 9.3\\
          & 2$_{0,2,1}$-1$_{0,1,1}$ & {E} & 96.7445 & 20.09 & 0.3 & 6$_{2,4,2}$-5$_{2,3,2}$ & {E} & 290.3073 & 74.66 & 9.5\\
          & 2$_{1,2,0}$-1$_{1,1,0}$ & {A+} & 95.9143 & 21.45 & 0.2 & 6$_{2,5,0}$-5$_{2,4,0}$ & {A-} & 290.1847 & 86.46 & 9.5\\
          & 2$_{1,1,1}$-1$_{1,0,1}$ & {E} & 96.7555 & 28.01 & 0.3 & 6$_{2,4,0}$-5$_{2,3,0}$ & {A+} & 290.2641 & 86.47 & 9.5\\
          & 5$_{1,5,2}$-4$_{0,4,1}$ & {E} & 84.5212 & 40.39 & 0.2 & 6$_{3,4,1}$-5$_{3,3,1}$ & {E} & 290.2132 & 96.47 & 8.0\\
          & 6$_{0,6,0}$-5$_{0,5,0}$ & {A+} & 290.1106 & 48.74 & 10.6 & 6$_{3,4,0}$-5$_{3,3,0}$ & {A+} & 290.1895 & 98.55 & 7.9\\
          & 3$_{2,1,0}$-4$_{1,4,0}$ & {A+} & 293.4641 & 51.64 & 2.9 & 6$_{3,3,0}$-5$_{3,2,0}$ & {A-} & 290.1905 & 98.55 & 7.9\\
          & 6$_{1,6,2}$-5$_{1,5,2}$ & {E} & 290.0697 & 54.32 & 10.3 & 6$_{3,3,2}$-5$_{3,2,2}$ & {E} & 290.2097 & 111.47 & 8.0\\
          & 6$_{0,6,1}$-5$_{0,5,1}$ & {E} & 289.9394 & 61.86 & 10.6& 6$_{4,3,0}$-5$_{4,2,0}$ & {A-} & 290.1613 & 129.10 & 5.9\\
          & 6$_{1,6,0}$-5$_{1,5,0}$ & {A+} & 287.6708 & 62.87 & 10.1 & 6$_{4,3,0}$-5$_{4,2,0}$ & {A+} & 290.1614 & 129.10 & 5.9\\
          & 6$_{1,5,0}$-5$_{1,4,0}$ & {A-} & 292.6729 & 63.71 & 10.6 & 6$_{4,3,2}$-5$_{4,2,2}$ & {E} & 290.1624 & 136.66 & 5.9\\
          & 6$_{1,5,1}$-5$_{1,4,1}$ & {E} & 290.2487 & 69.81 & 10.6 & 6$_{4,2,1}$-5$_{4,1,1}$ & {E} & 290.1833 & 144.76 & 5.9\\
          \hline
H$_2$CO     & 4$_{0,4}$-3$_{0,3}$ & {p} & 290.6234 & 34.90 & 69
           & 4$_{3,2}$-3$_{3,1}$ & {o} & 291.3804 & 140.94 & 30.4\\
            & 4$_{2,3}$-3$_{2,2}$ & {p}  & 291.2378 & 82.07 & 52.1
            & 4$_{3,1}$-3$_{3,0}$ & {o} & 291.3844 & 140.94 & 30.4\\
            & 4$_{2,2}$-3$_{2,1}$ & {p} & 291.9481 & 82.12 & 52.5
            & & & & & \\
\hline
\end{tabular}
\tablefoot{The line properties are taken from the CDMS and VASTEL catalogues. Limits were set for E$_{\mathrm{up}}$ $<$ 150 K (200 K for CH$_3$CCH) and A$_{\mathrm{ij}}$ $>$ 1.0$\times 10^{-6}$ (2.0 $\times 10^{-6}$ for CH$_3$OH).}
\end{table*}

\begin{figure*}[h!]
\centering
\includegraphics[width=\textwidth]{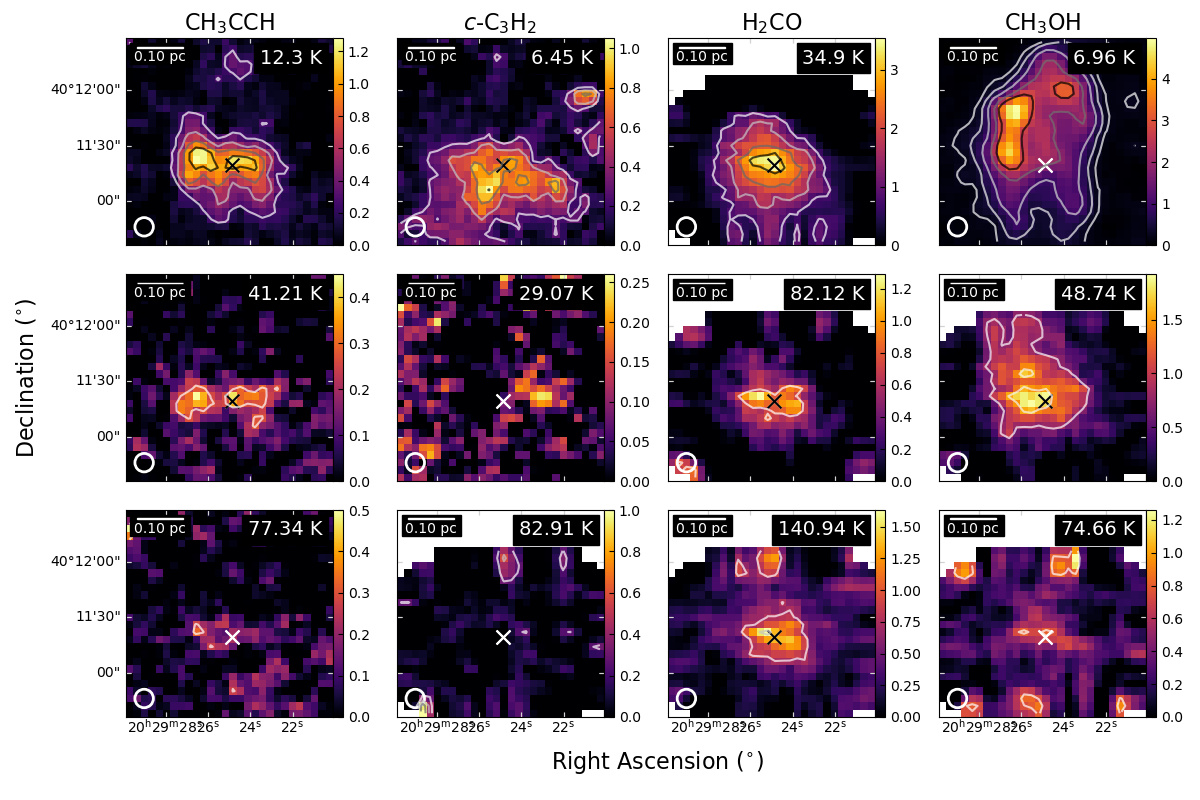}
\caption{AFGL 2591 integrated intensity $\int T_A\mathrm{d}v$ (K~km s$^{-1}$) maps of rotational transition lines spanning a range of energy levels. The upper state energy level is indicated in the top right-hand corner and the beam size in the bottom left-hand corner. The source VLA 3 is noted by a white or black `x'. CH$_3$CCH, in the first column, $c$-C$_3$H$_2$, in the second, H$_2$CO, in the third, have contour levels of 10, 20, 30, and 40 times the noise level of the respective frequency range. CH$_3$OH, in the fourth column, has contour levels of 10, 20, 40, 80, and 120 times the noise level of the respective frequency range.}
\label{fig:afgl2591map}
\end{figure*}

\begin{figure*}[h!]
\centering
\includegraphics[width=\textwidth]{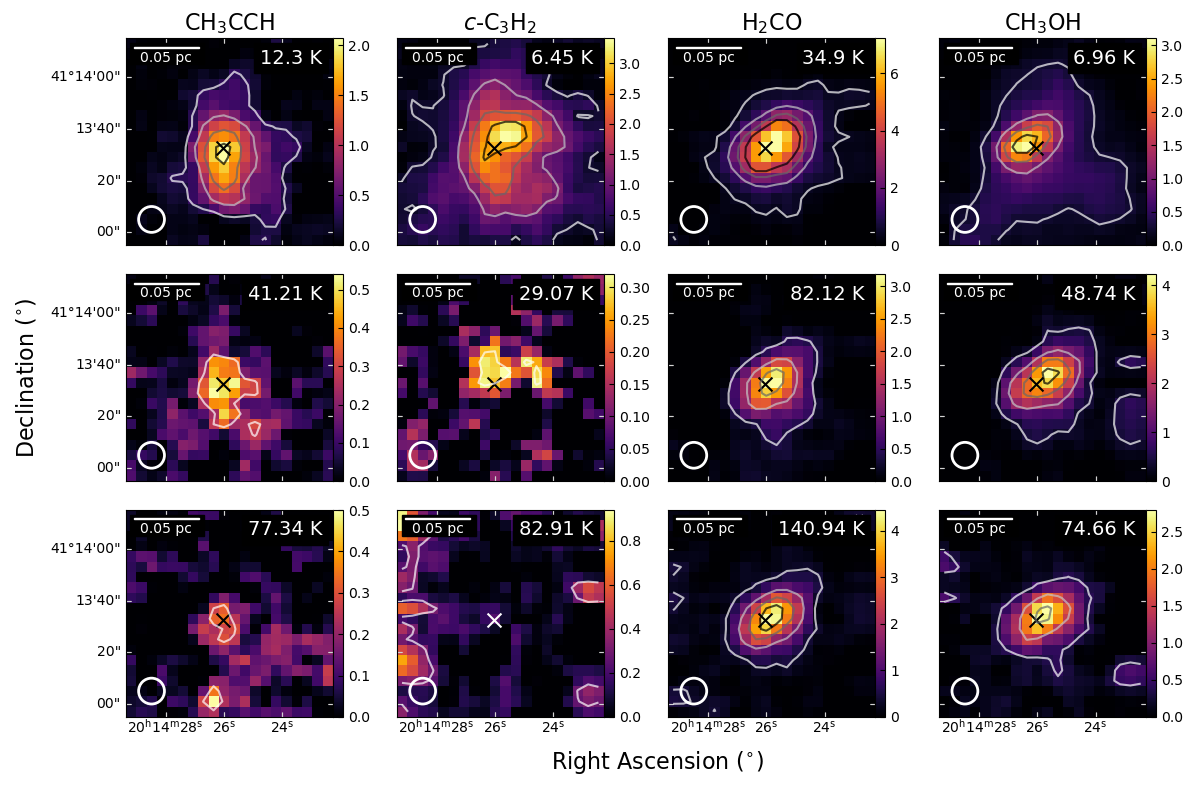}
\caption{Same as Fig.~\ref{fig:afgl2591map} for IRAS 20126, the source noted by a white or black `x'. CH$_3$CCH, in the first column, has contour levels of 10, 30, 50, and 70 times the noise of the respective frequency range. $c$-C$_3$H$_2$, in the second column, H$_2$CO, in the third, and CH$_3$OH, in the fourth, have contour levels of 10, 40, 70, and 100 times the noise of the respective frequency range.}
\label{fig:iras20126map}
\end{figure*}

Figures \ref{fig:afgl2591map} and \ref{fig:iras20126map} show integrated intensity maps for select lines covering a range of upper energy levels. These are notably bright in the lower energy lines, and decrease in strength as the upper state energy of the lines increase. Upwards of a few tens of K, there is little structure to be seen in the carbon chain molecules. In AFGL 2591 (Fig.~\ref{fig:afgl2591map}), there are differences in spatial distribution among the different molecules as well as between the different energy level lines. Especially in CH$_3$OH, the warm lines of E$_{\mathrm{up}}$\,=\,20-50~K trace the cloud near the known sources VLA 1-5, while the colder 6-20~K lines trace a region, known to be abundant in methanol (referred to as the `methanol plume' by \citealt{vanderwiel2011}), to the north and north-east. The similarity between molecules is in the quite extended emission around the hot cores region.

In IRAS 20126 all maps are quite featureless. The COM distributions, across energy levels, have a slight south-east--north-west tilt, and are all similarly concentrated around a single region, coinciding with the hot core. The CCM lines show a similar spatial concentration however with a slight south-north extension in the lines detected, as opposed to the south-east--north-west extension seen in CH$_3$OH. The integrated intensities of the carbon chain lines drop off in strength considerably from the lowest energy lines of 12~K and 6~K, for CH$_3$CCH and $c$-C$_3$H$_2$, respectively.

The beam size, 10.0$^{\prime\prime}$, as displayed in the integrated intensity maps, corresponds to 3.3$\times$10$^4$~AU (0.16~pc) in AFGL 2591 and 1.6$\times$10$^4$~AU (0.08~pc) in IRAS 20126 which is sufficient to resolve the extended molecular emission of tens of thousands of AU in each source (\citealt{cesaroni1999, vanderwiel2011}).

\bigskip

\subsection{LTE modelling}
\label{section:ltemodel}

\subsubsection{Model description and implementation}

We used a more comprehensive LTE model to determine the physical conditions of AFGL 2591 and IRAS 20126. The model, developed in Python, was based on the CASSIS LTE model scripts in Jython. This formalism is fully documented in Vastel's Formalism for the CASSIS software\footnote{http://cassis.irap.omp.eu/docs/RadiativeTransfer.pdf}. The LTE model simultaneously fit numerous transitions of one or more molecules across a spectrum for one or more physical components -- defined by the physical parameters source size, excitation temperature, column density, spectral line width, and source Doppler velocity -- of a source.

The model identified the molecular transitions producing spectral lines in a given frequency range. Then, it generated spectral line profiles for any desired molecule iterating over the physical parameters of the gas until it produced a model spectrum that best matches the data. The resulting best fit, therefore, provided these five physical parameters for the gas emitting the observed spectra. As we produced maps with our observations, the code was modified to loop over multiple pixels modelling all the observed spectra, ultimately producing spatial maps of the physical conditions.

The best fit was determined via the Levenberg-Marquardt algorithm in LMFIT \citep{newville2014}, a curve-fitting method for non-linear least-squares problems. The Levenberg-Marquardt algorithm is sufficiently quick for finding the minimum value, and we trust that with appropriately restricted parameter ranges (initial estimations are described below and adjusted in successive iterations of the model) we found the global minimum. The goodness-of-fit was reported with the reduced $\chi^2$, which describes the least-squares minimisation and takes into account the number of data points and variables included in the model. For an LTE model, there may be degeneracy between parameters such as column density and temperature. In some cases, we have few spectral lines compared to the number of free parameters and cannot exclude the possibility of degeneracy. However, we present a best effort to produce reasonable results given this limitation. In other cases, we have several strong lines or have fixed certain parameters in order to reduce the discrepancy between data points and free parameters. This is discussed in the LTE model results below.

In the analysis, we excluded the possible CH$_3$OH maser line at 84.5~GHz ($5_{1,5,2}-4_{0,4,1}$), seen in \citealt{kalenskii2002}, for example. While determining maser sources is outside the scope of this paper, our initial LTE models for identifying lines greatly under-represented the 84.5~GHz line strengths in several pixels with strong CH$_3$OH emission. Thus, we removed it in case it was a maser. Within our frequency ranges, no other methanol masers are reported in these sources.

We restricted the number of pixels the model loops through with a signal-to-noise ratio (S/N) mask for each molecule of interest. We fit a Gaussian to the spectrum using Astropy \citep{astropy}, calculated the peak of the observed spectrum using the Gaussian1D fitter and calculated the noise from a nearby, line-free frequency range with the SigmaClip function. We fit specific, strong lines of each molecule to create this mask. Any pixel that had a S/N $>$ 2.5 for the CH$_3$CCH 12~K 5$_0$-4$_0$ (E$_{\mathrm{up}}$\,=\,12~K) and 5$_1$-4$_1$ (E$_{\mathrm{up}}$\,=\,19~K) lines at 85~GHz are included in the CH$_3$CCH and $c$-C$_3$H$_2$ LTE models, and any pixel that had the same threshold for the CH$_3$OH 2$_{0,2,0}$-1$_{0,1,0}$ (E$_{\mathrm{up}}$\,=\,7~K) and 2$_{1,2,2}$-1$_{1,1,2}$ (E$_{\mathrm{up}}$\,=\,12~K) CH$_3$OH lines at 96~GHz or the 6$_{0,6,0}$-5$_{0,5,0}$ (E$_{\mathrm{up}}$\,=\,48~K) and 6$_{1,6,2}$-5$_{1,5,2}$ (E$_{\mathrm{up}}$\,=\,54~K) lines at 290~GHz (to capture lines in both GBT and IRAM data) are used for the CH$_3$OH and H$_2$CO models. The S/N threshold of 2.5 is lower than the standard 3 and was used to ensure we capture the cloud edges where, from a visual inspection of the spectral lines, we believed the emission is still real. As well, the corresponding $c$-C$_3$H$_2$ and H$_2$CO masks produced more extended maps. To allow comparison between molecules we continued with the more restrictive CH$_3$CCH and CH$_3$OH maps with a loosened S/N threshold.

To begin the analysis, we extracted the spectrum from the brightest pixel of certain molecular transition lines -- CH$_3$CCH 5$_0$-4$_0$ (E$_{\mathrm{up}}$\,=\,12~K); $c$-C$_3$H$_2$ 2$_{1,2}$-1$_{0,1}$ (E$_{\mathrm{up}}$\,=\,6~K); CH$_3$OH 2$_{0,2,0}$-1$_{0,1,0}$, 2$_{1,2,2}$-1$_{1,1,2}$, 2$_{0,2,1}$-1$_{0,1,1}$ triplet (E$_{\mathrm{up}}$\,=\,7, 12, 20~K), CH$_3$OH 6$_{0,6,0}$-5$_{0,5,0}$ (E$_{\mathrm{up}}$\,=\,48~K); H$_2$CO 4$_{0,4}$-3$_{0,3}$ (E$_{\mathrm{up}}$\,=\,35~K) -- and fit one or more Gaussians to these lines with the CASSIS Line Analysis module. This module, with a chosen molecule, frequency range, and transition thresholds (such as $E_{\rm up, max}$ and $A_{\rm ij, min}$), selected and displayed all possible transition lines in the data. By fitting Gaussians by hand, we determined the number of possible components to use in the LTE model and used the line width and velocity as guides for the input parameters. The temperature and column density, on the other hand, are physical properties of the source and input ranges are estimated based on previous knowledge of these sources as well as our integrated intensity maps, Figs.~\ref{fig:afgl2591map}, and~\ref{fig:iras20126map}. The size of the source was set at 20$\arcsec$, assumed to be two times the size of the beam. This was to account for the extended emission seen by single dish telescopes, and it was appropriate as we are not resolving the compact protostellar sources but the larger scale envelope. Assuming a source size larger than the beam also removed any beam dilution effects when modelling the molecular spectral lines. With smaller (10$\arcsec$) beam sizes, the resulting column density changed but remained within a factor of 3. With larger (30$\arcsec$) beam sizes, the column density changed by less than a factor of 2. Otherwise, the results did not change and we maintained this constant beam size.

\subsubsection{AFGL 2591 LTE model results}
\label{sec:afgl2591ltemodel}

Initially, each molecule was modelled individually with one component, unless the CASSIS Gaussian fitting described suggested there were multiple components -- either through different velocity components, line wings, or otherwise non-Gaussian shape. If multiple components were needed, certain values were fixed to balance the number of transition lines available and the number of free parameters in the model. The input parameters for all models are shown in Table~\ref{tab:ltemodelinput}.

The initial one-component fit of $c$-C$_3$H$_2$ alone provided unrealistic excitation temperatures (minimising at the lower limit of 3~K) and corresponding large column densities that did not align with the observed emission. There is only weak detection of the lowest energy (E$_{\mathrm{up}}$\,=\,6~K) $c$-C$_3$H$_2$ transition line which dominates the fits. Given this, we were unable to model with $c$-C$_3$H$_2$ alone and it was modelled in the same component as CH$_3$CCH in the CH$_3$CCH one-component model. We assumed the two carbon chain species are originating from the same gas in the extended warm and cold extended envelope. The additional transition lines provided further constraints on the resulting excitation temperature and column density from the model. When combined with CH$_3$CCH, $c$-C$_3$H$_2$ reproduces the same structure of the integrated intensity maps of $c$-C$_3$H$_2$ in the column density maps, indicating that the column density reflects where we see emission in the data. Previous models, with $c$-C$_3$H$_2$ alone, produced unrealistic column density distributions. In addition, the model fits the spectral lines for each molecule well (example in Fig.~\ref{fig:iras20126ccmresults}), supporting the assumption that these two molecules are coming from the same velocity and temperature gas. In further support, these are different conditions from the other two molecules, described next. As will be described in the following section for the results of IRAS 20126, $c$-C$_3$H$_2$ could also show some non-LTE behaviour.

The CH$_3$OH spectral line profiles contain more velocity structure than compared to the other molecules. CH$_3$OH was fit unsuccessfully by a one-component fit, as determined by non-converging models, high reduced $\chi^2$ values, or unphysical values for the excitation temperature or column density. The low energy GBT lines were often over-represented compared to the higher energy IRAM lines, and produced very low excitation temperatures unexpected for LTE. 

\begin{figure}[h!]
\centering
\includegraphics[trim=2.2cm 0.8cm 2.3cm 2cm, clip=true, width=0.5\textwidth]{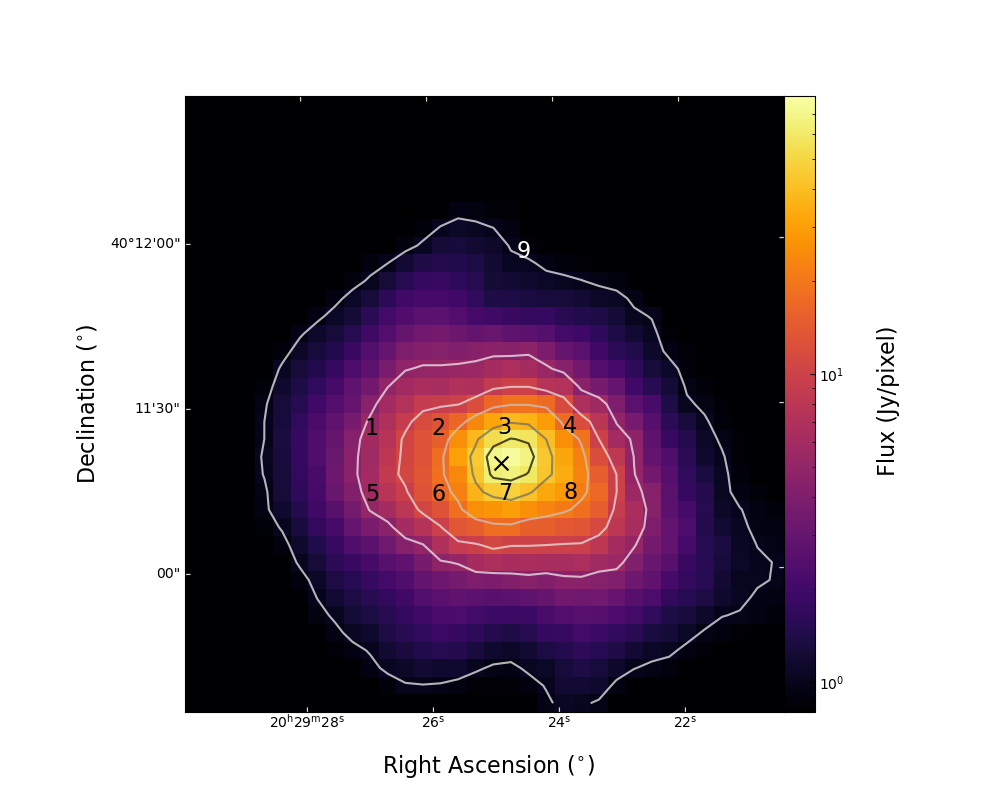}
\caption{Pixels fit with a RADEX model in AFGL 2591 marked on a Herschel PACS continuum image. The black and white numbers correspond to said pixels, with results listed in Table~\ref{tab:radex}. The black `x' represents the source VLA 3.}
\label{fig:afgl2591radex}
\end{figure}

\begin{figure}[h!]
\centering
\includegraphics[scale=0.50]{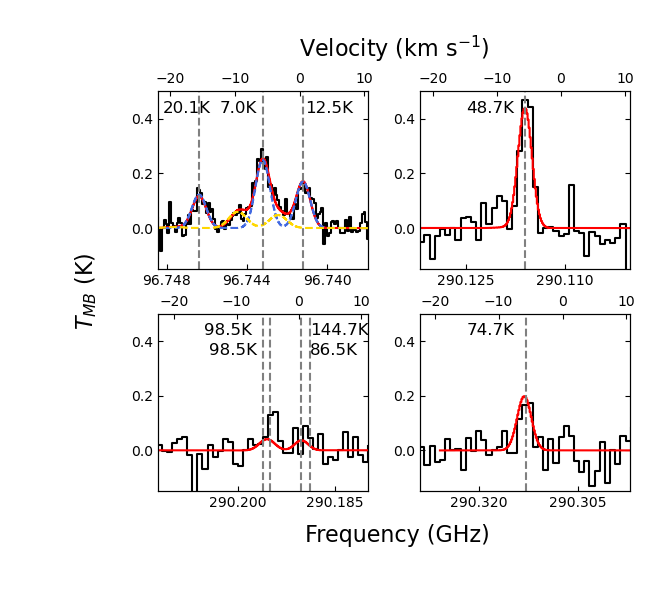}
\caption{RADEX fit for CH$_3$OH in the central pixel of AFGL 2591, corresponding to pixel 7 in Fig.~\ref{fig:afgl2591radex} and Table~\ref{tab:radex}. The solid red line is the overall two-component model, the solid black line is the data, the dashed blue line is component one and the dashed yellow line is component two. Component two is only seen in the lower energy lines within the GBT range, thus component one and the overall model are the same for the lines in the IRAM range. The dashed grey line represents the observed transition frequency, with the energy level of this transition labelled.}
\label{fig:afgl2591ch3ohradex}
\end{figure}

Given the possible non-LTE nature as indicated by the low excitation temperatures found, we tried to model CH$_3$OH in RADEX \citep{vandertak2007} with CASSIS. Unlike the LTE model, RADEX is too intensive to run over all pixels in the map. Thus, select points near the source VLA 3 and along the eastward extension of the molecules are modelled, along with one pixel off-source to the north, as seen in Fig.~\ref{fig:afgl2591radex}. Table~\ref{tab:radex} shows the results, where the model delineates the source into two components -- one which traces the higher column density (10$^{14}$~cm$^{-2}$), warmer (20~K), gas near the source v$_{lsr}$ of $-$5.7~km s$^{-1}$ represented by the higher energy IRAM lines, and a second which traces the colder (10~K), lower column density (10$^{13}$~cm$^{-2}$) more negative velocity ($-$7 to $-$9~km s$^{-1}$) gas. Figure~\ref{fig:afgl2591ch3ohradex} shows the resulting model spectra for pixel (38,31).

The kinetic temperature from the RADEX model is then used to guide the input excitation temperature, a key parameter for this molecule, for the LTE model. Since the kinetic and excitation temperatures matched, the assumption of LTE was valid and we used the LTE model to produce parameter maps for CH$_3$OH. In subsequent LTE modelling, the latter cold component was able to be differentiated into two components in the LTE model -- the CH$_3$OH `blob' towards the north-east below in component two and the blue-shifted component three, both described below.

H$_2$CO, lastly, is well represented by a one-component model where the parameters, aside from size, are free. Taking direction from the CCMs, we tried to fit H$_2$CO and CH$_3$OH in the same component of the model to represent these molecules originating from the same gas. However, none of the multiple colder components of CH$_3$OH aligned with the hotter H$_2$CO emission.

\begin{figure*}[h!]
\centering
\includegraphics[width=\textwidth]{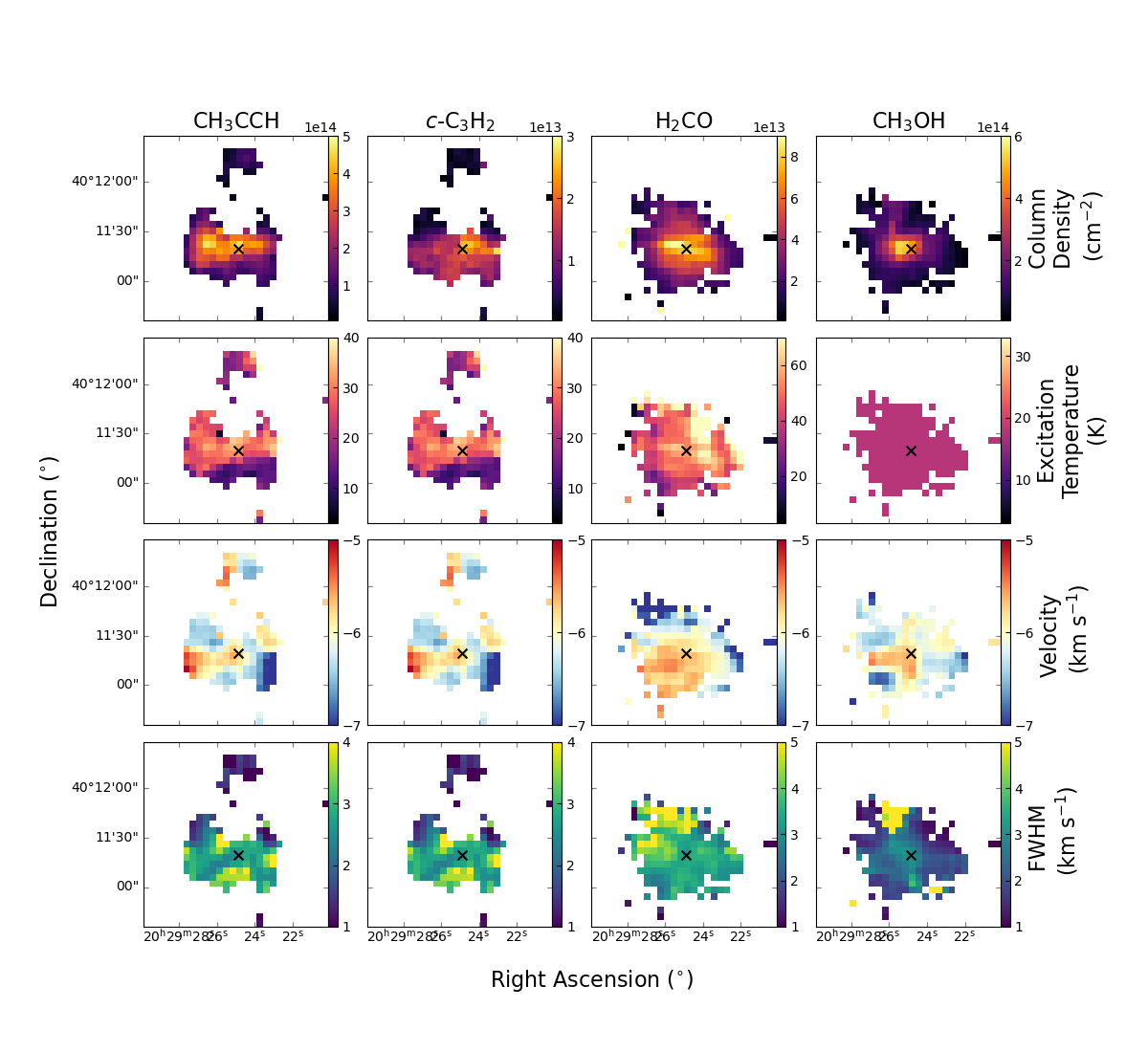}
\caption{Parameters maps of AFGL 2951 showing the results of the LTE models. The black `x' represents the source VLA 3. For CH$_3$OH, T$_{ex}$ was fixed at 18~K.}
\label{fig:afgl2591paramresults}
\end{figure*}

The parameter maps, for velocity, excitation temperature, column density, and line width (FWHM) of the four molecules are displayed in Fig.~\ref{fig:afgl2591paramresults}. For all parameter maps, the first component is shown for each molecule. This component may also be seen as the `main' component, as it is most closely associated with the protostar and the envelope. The second and third components for CH$_3$OH are displayed in Fig.~\ref{fig:afgl2591paramresults2}.

All four molecules show a fairly constant velocity structure near the source v$_{lsr}$ of $-$5.7 ($\pm$0.03-0.5)~km s$^{-1}$. Component three of CH$_3$OH, however, is notably blue-shifted and fixed at $-$9.4~km s$^{-1}$. Component two of CH$_3$OH, representing the offset methanol `blob,' has no contribution towards the centre leaving a hole in the map. Towards the edges, where the detected lines become weaker, this value is slightly blue-shifted.

The carbon chain molecules show a fairly uniform temperature of 30 ($\pm$2-4)~K across the main emitting areas of the source. The excitation temperature drops to near 10~K where the all lines are weaker and the $c$-C$_3$H$_2$ lines dominate. The temperature of H$_2$CO reaches towards 70 ($\pm$5)~K slightly offset from the VLA 1-3 sources and shows a warm temperature across extended spatial scales. For CH$_3$OH, not displayed, the excitation temperatures were fixed at 18~K for component one and 8~K for components two and three.

For all components, the column density maps trace features of the integrated intensity maps in Fig.~\ref{fig:afgl2591map}. CH$_3$CCH, at a peak of 4 ($\pm$0.4) $\times$10$^{14}$~cm$^{-2}$, extends along and towards the east of AFGL 2591, with two distinct peaks. $c$-C$_3$H$_2$ has the lowest column density, 2 ($\pm$0.3) $\times$10$^{13}$~cm$^{-2}$, and shows an extended distribution with a slight peak near the VLA sources. H$_2$CO peaks at 9($\pm$0.03) $\times$10$^{13}$~cm$^{-2}$, and distinctly is distributed near the VLA sources. CH$_3$OH peaks at 5 ($\pm$0.2), 5 ($\pm$0.2), and 2 ($\pm$0.1) $\times$10$^{14}$~cm$^{-2}$ for components one, two, and three, respectively. Component one has the highest column density of all molecules and peaks slightly offset of the protostars, similar to the other molecules. Component two shows the extended emission to the north-east found in the lower energy transition lines, while component three shows a less dense, blue-shifted component just north of the source.

The line widths of the carbon chain molecules are around 3~km s$^{-1}$ wide across the emitting region, and H$_2$CO is uniform at a slightly broader width near 4~km s$^{-1}$. Components two and three of CH$_3$OH were fixed at 3.9~km s$^{-1}$ and 2.5~km s$^{-1}$ respectively, while component one results in line widths of 2-3~km s$^{-1}$.

While all our detected lines represent extended envelope emission, the differing temperatures in LTE suggest the molecules are in different gas. As well, the multiple components reveal different structures in the region, which is examined in Sect.~\ref{section:afgl2591discussion}. \\

\begin{table*}
\centering
\caption{RADEX results for CH$_3$OH in AFGL 2591.}
\label{tab:radex}
\begin{tabular}{ l c c c c c }
\hline\hline
\multicolumn{1}{c}{Pixel} & \multicolumn{1}{c}{\text{T$_{kin}$}}  & \multicolumn{1}{c}{\text{V$_{\mathrm{lsr}}$}} & \multicolumn{1}{c}{\text{N$_{\mathrm{tot}}$}}  & \multicolumn{1}{c}{\text{N$_{\mathrm{H2}}$}}  & \multicolumn{1}{c}{\text{FWHM}} \\
\multicolumn{1}{c}{} & \multicolumn{1}{c}{\text{(K)}} &  \multicolumn{1}{c}{\text{(km s$^{-1}$)}}& \multicolumn{1}{c}{\text{($\mathrm{cm^{-2}}$)}} & \multicolumn{1}{c}{\text{($\mathrm{cm^{-3}}$)}} & \multicolumn{1}{c}{\text{(km s$^{-1}$)}}\\
\hline
1 (32,34) & 12.1 & -6.15 & 4.1$\times10^{14}$ & 6.4$\times10^{7}$ & 2.8 \\
        & 11.0 & -7.30 & 12.9$\times10^{13}$ & 1.0$\times10^{5}$ & 3.2 \\
2 (35,34) & 17.5 & -6.35 & 4.1$\times10^{14}$ & 1.5$\times10^{7}$ & 3.1 \\
        & 10.6 & -9.50 & 6.7$\times10^{13}$ & 1.0$\times10^{5}$ & 3.2 \\
3 (38,34) & 19.5 & -5.95 & 3.5$\times10^{14}$ & 3.9$\times10^{8}$ & 2.8 \\
        & 11.2 & -9.50 & 10.0$\times10^{13}$ & 1.0$\times10^{5}$ & 3.2 \\
4 (41,34) & 16.5 & -6.00 & 2.3$\times10^{14}$ & 1.0$\times10^{7}$ & 1.9 \\
        & 10.9 & -9.30 & 3.2$\times10^{13}$ & 1.0$\times10^{5}$ & 3.2 \\
5 (32,31) & 14.5 & -5.30 & 2.4$\times10^{14}$ & 3.8$\times10^{6}$ & 3.1 \\
        & 10.2 & -7.10 & 3.1$\times10^{13}$ & 1.0$\times10^{5}$ & 3.2 \\
6 (35,31) & 16.5 & -5.70 & 3.6$\times10^{14}$ & 7.8$\times10^{6}$ & 2.6 \\
        & 10.8 & -8.90 & 14.9$\times10^{13}$ & 1.0$\times10^{5}$ & 3.2 \\
7 (38,31) & 21.4 & -5.75 & 2.6$\times10^{14}$ & 3.8$\times10^{7}$ & 2.6 \\
        & 10.3 & -9.60 & 2.6$\times10^{13}$ & 1.0$\times10^{5}$ & 3.2 \\
8 (41,31) & 19.1 & -6.15 & 2.0$6\times10^{14}$ & 1.1$\times10^{7}$ & 2.4 \\
        & 10.3 & -9.00 & 3.4$\times10^{13}$ & 1.5$\times10^{5}$ & 3.2 \\
9 (39,42) & 15.3 & -4.85 & 6.6$\times10^{13}$ & 1.9$\times10^{6}$ & 3.2 \\
        & 11.6 & -8.8 & 19.2$\times10^{13}$ & 1.0$\times10^{5}$ & 3.2 \\
\hline
\end{tabular}
\tablefoot{Eight pixels in E-W strips across the central source of AFGL 2591, and one off-source, were taken and fit with RADEX, a non-LTE model. These are labelled in black and white in Fig.~\ref{fig:afgl2591radex}. Two components were fit to each pixel's spectra. A- and E-CH$_3$OH are used in this model due to the availability of collisional co-efficients. There are not enough lines to determine the A-/E- ratio, thus it is assumed to be 1.}
\end{table*}

\subsubsection{IRAS 20126 LTE model results}
\label{sec:iras20126ltemodel}

\begin{figure*}[h!]
\centering
\includegraphics[width=\textwidth]{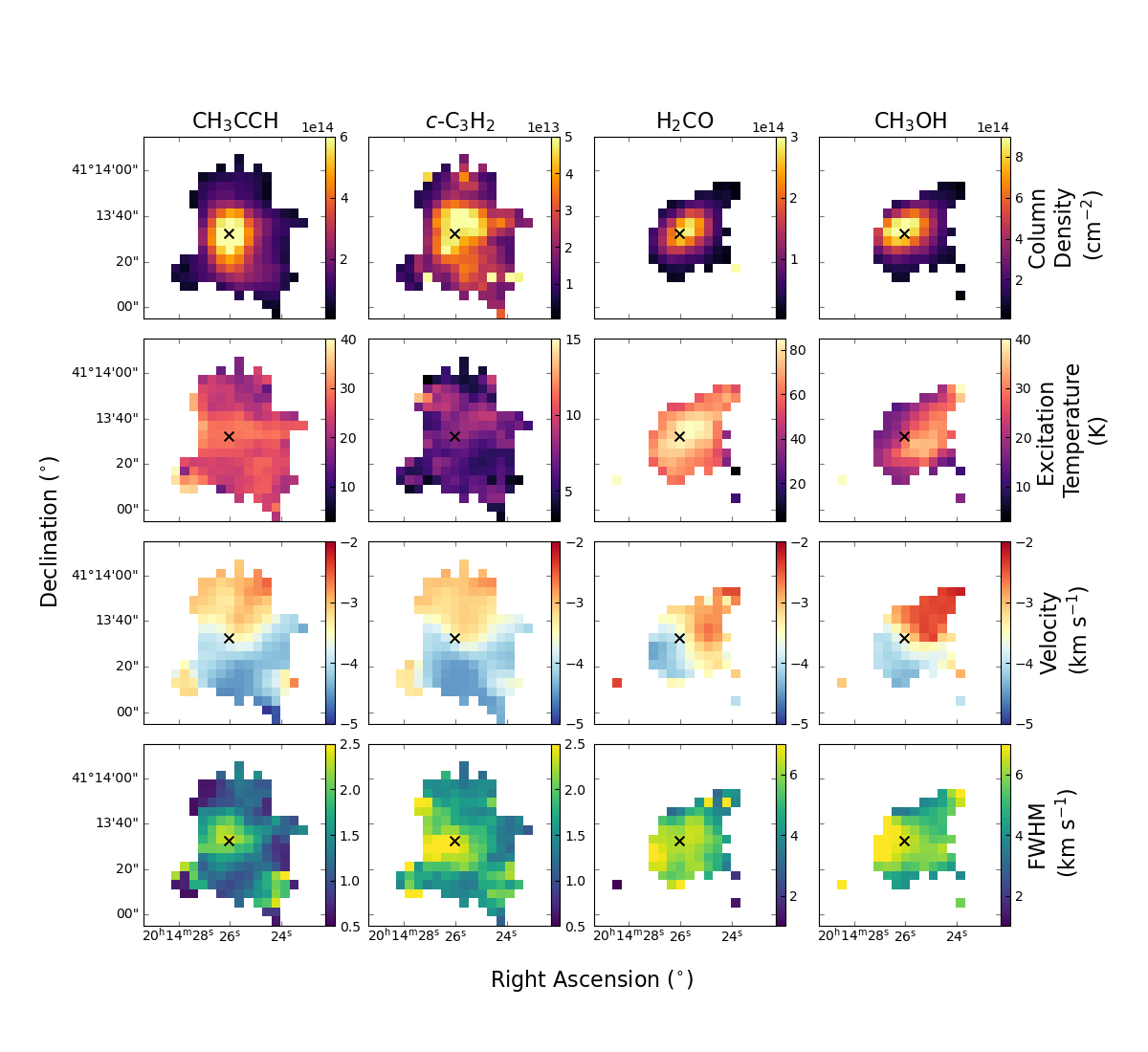}
\caption{Parameters maps of IRAS 20126 showing the results of the LTE models. The black `x' represents the protostellar source.}
\label{fig:iras20126paramresults}
\end{figure*}

As in AFGL 2591, each molecule is initially modelled individually. For IRAS 20126, this produced the best results and the presented analysis keeps all molecular models separate. Contrary to AFGL 2591, the $c$-C$_3$H$_2$ and CH$_3$CCH combination did not represent the lines well. The $c$-C$_3$H$_2$ lines in IRAS 20126 are stronger, and drove the excitation temperatures too low for CH$_3$CCH. CH$_3$OH and H$_2$CO were also kept separate, as the slight disparities in temperature worsened the fits if combined. Given the limitations of the number of components we could discern with this data and this modelling, there may be shared components which we do not see.

Each molecule is modelled with one component unless the CASSIS Gaussian fit suggested there were multiple components. Figure~\ref{fig:iras20126spectra} displays the model results for select transition lines covering a range of upper energy levels. For CH$_3$CCH and H$_2$CO, one component was adequate. For $c$-C$_3$H$_2$ and CH$_3$OH, a second narrow, cold component was added to better represent the detected lines. This component, in both molecules, is a foreground screen across the source in the lower energy transition lines -- as seen in the $c$-C$_3$H$_2$ 2$_{1,2}$-1$_{0,1}$ (E$_{\mathrm{up}}$\,=\,6~K) and CH$_3$OH 2$_{0,2,0}$-1$_{0,1,0}$ (E$_{\mathrm{up}}$\,=\,7~K) lines in  Fig.~\ref{fig:iras20126spectra}. In these cases of multiple components, certain values were fixed to balance the number of transition lines available and the number of free parameters in the model. In giving the errors of the varying components, we present the statistical errors from the model in the range seen across the map, excluding edges.

The parameter maps, for velocity, excitation temperature, column density, and line width of the four molecules are displayed in Fig.~\ref{fig:iras20126paramresults}. For all parameter maps, the main or first component is shown for each molecule. The second components of $c$-C$_3$H$_2$ and CH$_3$OH are displayed in Fig.~\ref{fig:iras20126paramresults2}.

There is a consistent velocity range for all species between $-$2.0 to $-$4.5 ($\pm$0.02-0.1)~km s$^{-1}$. This is a range around the source V$_{lsr}$ of $-$3.5~km s$^{-1}$. For the CCMs this is oriented with a red-shifted northern and blue-shifted southern feature. For the COMs, this is oriented with a red-shifted north-west and blue-shifted south-east feature. In $c$-C$_3$H$_2$, the second component was fixed at the same velocity as the first component, and in CH$_3$OH the second component was fixed at $-$0.2~km s$^{-1}$\,from the first.

Similar to the velocity, the excitation temperature structure is differentiated between categories of molecules. The carbon chains have a roughly consistent temperature across the source, with CH$_3$CCH at 25-30 ($\pm$0.9-4)~K and $c$-C$_3$H$_2$ at 8-12 ($\pm$0.8-3)~K. The COMs show a warmer central region falling off in temperature towards the edges, with a range of 15-35 ($\pm$0.3-2)~K for CH$_3$OH located just off the protostar, and 55-85 ($\pm$3-6)~K for H$_2$CO located on the protostar. The second component of $c$-C$_3$H$_2$ varies from 4-8 ($\pm$0.4-6)~K with few pixels at 10-12~K, while the second component of CH$_3$OH was fixed at 10~K.

\begin{figure}[h!]
\centering
\includegraphics[scale=0.50]{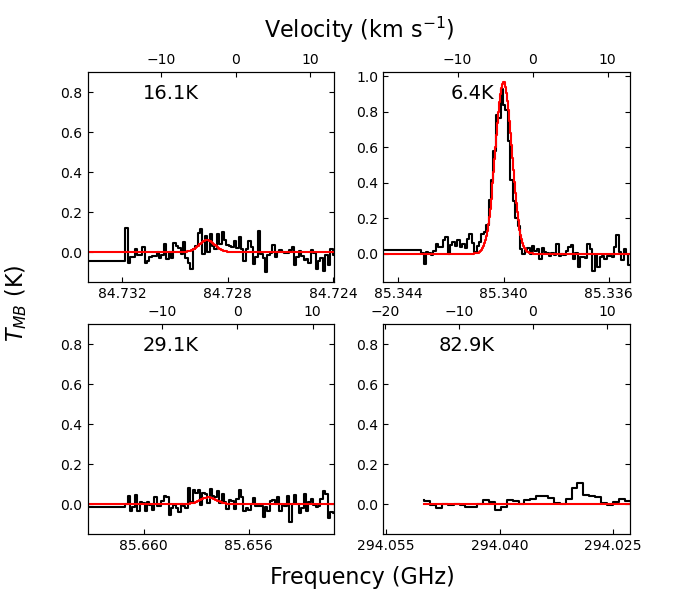}
\caption{RADEX fit for $c$-C$_3$H$_2$ in the central pixel of IRAS 20126. The model used a temperature of 30~K, as expected from the CH$_3$CCH results, an ortho- to para- ratio of 3, and a density of 10$^5$~cm$^{-3}$. The model is in red while the data is in black. The upper state energy level of each transition is shown on each plot.}
\label{fig:iras20126c3h2radex}
\end{figure}

Given the low excitation temperature of $c$-C$_3$H$_2$, it is possible that it is not in LTE and these results may not be a true representation of the physical gas temperature. To explore this, we used RADEX to model a pixel that was bright in $c$-C$_3$H$_2$, as shown in Fig.~\ref{fig:iras20126c3h2radex}. We reproduced the spectrum using a temperature of 30~K, as expected from the CH$_3$CCH results, an ortho- to para- ratio of 3, and a density of 10$^5$~cm$^{-3}$.

All column density maps peak near the site of the protostar, with the COMs showing a steeper falloff towards the edges. CH$_3$CCH, CH$_3$OH, and H$_2$CO peak near 6, 9, and 3 ($\pm$ 0.3, 0.2, and 0.1) $\times$10$^{14}$~cm$^{-2}$ respectively, while $c$-C$_3$H$_2$ peaks near 5 ($\pm$0.3) $\times$10$^{13}$~cm$^{-2}$. The second component of $c$-C$_3$H$_2$ was fixed at 1$\times$10$^{13}$~cm$^{-2}$ and for CH$_3$OH it was fixed at 2.1$\times$10$^{13}$~cm$^{-2}$.

The CCM lines are consistently narrower in line width, ranging from about 1-2.5~km s$^{-1}$, while the COM lines are 4-7km s$^{-1}$ wide. The former show a slightly wider line profile near the protostar, while the latter show a widening towards the south-east of the source. The second component of $c$-C$_3$H$_2$ was fixed at a line width of 0.5~km s$^{-1}$ while for CH$_3$OH it was fixed at 1.2~km s$^{-1}$.

The visual and quantitative differences in the results, namely velocity and temperature, between the types of molecules suggests they are coming from different physical environments. We elaborate on the known and observed structure in Sect.~\ref{section:iras20126discussion}.

\section{Discussion}
\label{section:discussion}

Detections of carbon chains in high-mass star-forming regions have increased over recent years, with candidate, but not confirmed, WCCC sources (\citealt{saul2015, taniguchi2018a}). \citet{mookerjea2012} present the first trace of WCCC in a high-mass star-forming region with detections of the simple chain C$_3$ in DR21(OH), another source with hot cores in Cygnus X. The C$_3$ abundance is consistent with the warm-up model of WCCC in \citet{sakai2008}. The emission and the resulting temperatures of 30-50~K -- a `lukewarm corino' -- are consistent with the warm envelope around the hot core. The extended spatial distribution and the low, but warm, temperatures are also seen in our sources AFGL 2591 and IRAS 20126.

Comparing carbon chain molecules to CH$_3$OH is a recently available and practised method of analysing star-forming regions through molecular observations. Table~\ref{tab:ntotratio} summarises several studies that used either CH$_3$CCH or $c$-C$_3$H$_2$, as in this study. At first glance, there is a spread of N(CH$_3$CCH)/N(CH$_3$OH), possibly due to the variety of studies and sources that examine this ratio. All CH$_3$CCH studies presented used single dish observations and resolved objects on scales of 1-7$\times$10$^4$ AU, comparable to this study, and consistent with this species being found in the cold and warm extended envelopes. In \citet{fayolle2015}, CH$_3$CCH is only detected with the IRAM 30m and not with higher resolution Submillimeter Array observations. There is also variation within the protostellar age -- \citet{fayolle2015} note their sources, showing N(CH$_3$CCH)/N(CH$_3$OH) $>$ 1, are potentially younger than typical hot core sources, while \citet{taniguchi2018b} find N(CH$_3$CCH)/N(CH$_3$OH) $>$ 1 for their source with a UCH\,{\footnotesize II} region. For $c$-C$_3$H$_2$, \citet{higuchi2018} look at scales of a few thousand AU, a smaller region, yet they find variation within the survey of protostars and comparable values to our sources.

\begin{figure*}
     \centering
     \begin{subfigure}[b]{\textwidth}
         \centering
         \includegraphics[width=\textwidth]{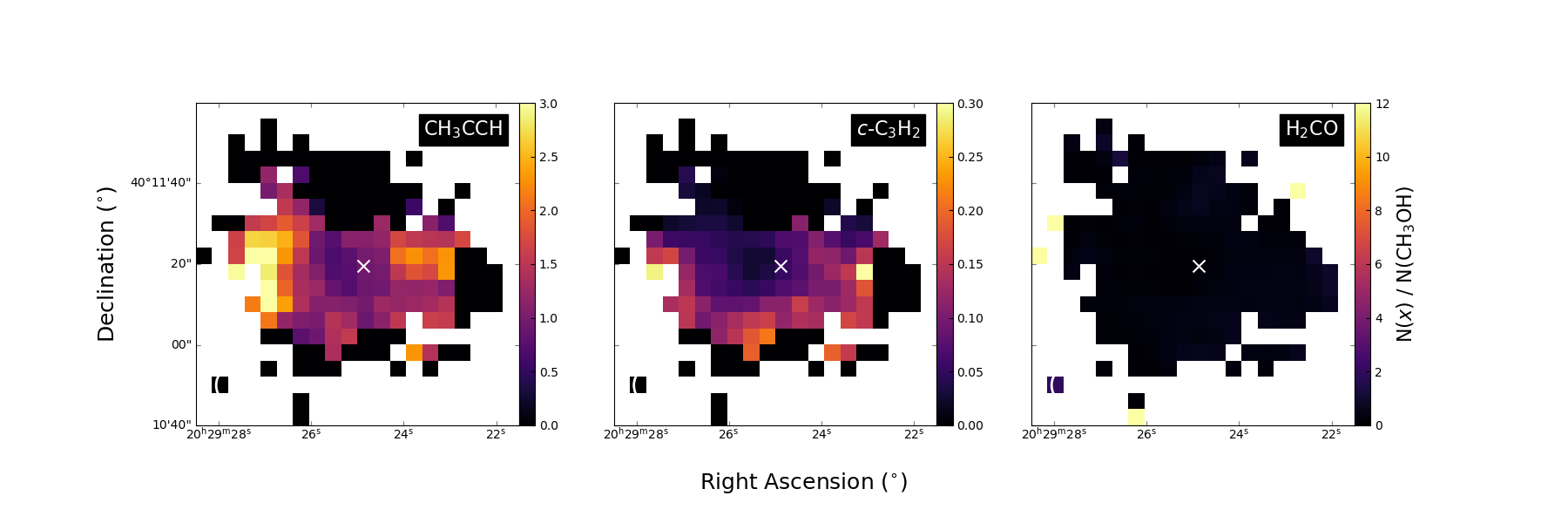}
     \end{subfigure}
     \hfill
     \begin{subfigure}[b]{\textwidth}
         \centering
         \includegraphics[width=\textwidth]{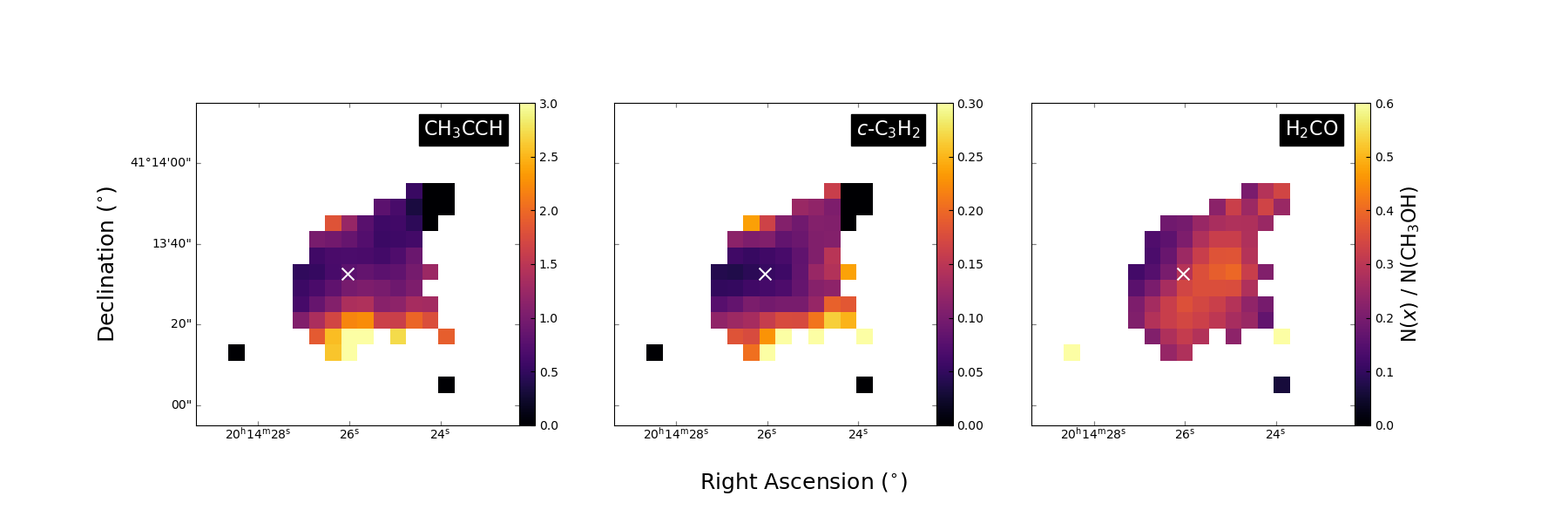}
     \end{subfigure}
        \caption{Column density ratio maps for CH$_3$CCH, $c$-C$_3$H$_2$, and H$_2$CO with respect to CH$_3$OH. The upper panel shows AFGL 2591 and the lower panel IRAS 20126. The white `x' represents the protostellar sources within each region.}
        \label{fig:ntotratio}
\end{figure*}

In this work, at the continuum peak of IRAS 20126 N(CH$_3$CCH)/N(CH$_3$OH) = 0.99 while N($c$-C$_3$H$_2$)/N(CH$_3$OH) = 0.06. For AFGL 2591, N(CH$_3$CCH)/N(CH$_3$OH) = 0.97 while N($c$-C$_3$H$_2$)/N(CH$_3$OH) = 0.05. The column density ratios are shown across the source maps in Fig.~\ref{fig:ntotratio}. The ratio maps in IRAS 20126 are fairly uniform and structured compared to AFGL 2591, which could arise from the smaller spatial extension seen in IRAS 20126, at half the distance of AFGL 2591. However, there is a wide range of ratios within a single source and not just from source to source, making cloud to cloud comparisons somewhat unreliable.

\begin{table*}
\caption{Column density ratios from the literature of carbon chain molecules with respect to methanol.}
\label{tab:ntotratio}
\centering
\begin{tabular}{ l l c c }
\hline\hline
Source & Source Type & N(CH$_3$CCH)/N(CH$_3$OH) & N($c$-C$_3$H$_2$)/N(CH$_3$OH)\\ 
\hline
G331.51-0.103$^a$ & hot molecular core & 0.42$\pm$0.05 & - \\
G12.89+0.49$^b$ & massive young stellar object & 0.34$^{+0.28}_{-0.15}$ &  -  \\
G16.86-2.16$^b$ & massive young stellar object & 0.36$^{+0.28}_{-0.18}$ &  - \\
G28.28-0.36$^b$ & massive young stellar object & 1.61$^{+1.4}_{-0.86}$ & - \\
Inner galaxy $^c$ & ATLASGAL high mass clumps, averaged & 0.31 &  - \\
NGC 7538 IRS 9$^d$ & organic-poor massive young stellar object & 1.3$\pm$0.4 &  - \\
W3 IRS5$^d$ & organic-poor massive young stellar object & 2.2$\pm$0.7 & - \\
AFGL 490$^d$ & organic-poor massive young stellar object & 1.8$\pm$0.8 & - \\
Perseus molecular cloud $^e$ & Class 0/1 protostars & -  & 0.009$\pm$0.003-0.80$\pm$0.34 \\
L1527$^e$ & low-mass star-forming region & - & 0.60$\pm$0.07 \\
IRAS 20126$^f$ & high-mass star-forming region & 0.99 & 0.06 \\
AFGL 2591$^f$ & high-mass star-forming region & 0.97 & 0.05 \\
\hline
\end{tabular}
\tablebib{$^a$\citet{santos2022}, $^b$\citet{taniguchi2018b}, $^c$\citet{gianetti2017}, as averaged in $^a$, $^d$\citet{fayolle2015}, $^e$\citet{higuchi2018} and references therein, $^f$this work.}
\end{table*}

Environmental differences cause the variation between CCMs and CH$_3$OH. In our data, the column densities of CH$_3$CCH and CH$_3$OH are of comparable magnitude, and CH$_3$CCH is typically seen on a uniform, extended scale in comparison to concentrated CH$_3$OH closer to protostellar sources. \citet{santos2022} most recently compare N(CH$_3$CCH)/N(CH$_3$OH) in the massive hot molecular core G331.512-0.103 with single-point APEX observations and find a ratio of 0.42. The authors compare abundance ratios to that of other high-mass star-forming regions, with values across a range of 0.31-2.2 (with error, \citealt{taniguchi2018b, gianetti2017, fayolle2015}). CH$_3$OH is likely to be more spatially compact than CH$_3$CCH, however. The latter is typically an envelope species, as in \citet{fayolle2015} where the ratios are $>$1. This study focuses on organic poor massive young stellar objects, rather than CH$_3$OH rich hot cores, though the presence and relative abundance of CH$_3$CCH and other complex species makes them comparable to, or possible precursors to, high mass hot cores.

$c$-C$_3$H$_2$ is often an order of magnitude lower in column density than CH$_3$OH, with no correlation between the two molecules. This is not surprising, due to the different formation routes but has been discussed in previous observations. \citet{spezzano2016} find different environmental conditions for $c$-C$_3$H$_2$ and CH$_3$OH in the prestellar core L1544, with interstellar radiation keeping C in its atomic form and allowing carbon chains to form near the edge of the source -- a new view of carbon chain chemistry. In the shielded region of the cloud, CO forms, leading to the production of complex organic molecules. Thus, there is a stark difference in spatial distribution between the two molecules (their Fig. A.1). L1554, however, is a much different environment than our two protostellar regions. In general, as a collapsing core evolves, the gas and dust environment will change. The original view of WCCC, which will be discussed for our sources in Sect.~\ref{chemicalmodel}, is dependent upon the timescale, where carbon chains are reproduced in the warm protostellar environment. More specifically, AFGL 2591 and IRAS 20126 are relatively isolated sources and have no external irradiation sources to affect chemical processes. For a more direct comparison, \citet{higuchi2018} find no correlation between $c$-C$_3$H$_2$ and CH$_3$OH for 36 protostars in the Perseus region. The column density ratio they find is on order of magnitude 10$^{-2}$-10$^{-1}$, similar to what we find in both IRAS 20126 and AFGL 2591.

\subsection{AFGL 2591}
\label{section:afgl2591discussion}

The environment of AFGL 2591 has well-known features: sources VLA 1-5, a warm inner envelope, and an extended methanol plume (\citealt{Hasegawa1995, vanderwiel2011, jimenezserra2012, gieser2019}). An extended distribution around this region is also seen in smaller molecules such as CO, HCN, HCO+, H$_2$CO \citep{vandertak1999}. Traces of these features are able to be seen from our LTE model (Fig.~\ref{fig:afgl2591paramresults}). The sources VLA 1-5 are not spatially resolved in our single dish observations, and the structures present in both our integrated intensity maps (Fig.~\ref{fig:afgl2591map}) and column density maps (Fig.~\ref{fig:afgl2591paramresults}) -- aside from the CH$_3$OH 2$_{0,2,0}$-1$_{0,1,0}$ (E$_{\mathrm{up}}$\,=\,7~K) line and component two CH$_3$OH maps, which trace the methanol `plume' -- are showing the smoothed out sources as well as the inner envelope. The dense inner envelope is shown to be a dominant component for transition lines with E$_{\mathrm{up}}\leq$\,200 K \citep{vanderwiel2013}.

VLA 3, an early B-type star and the central heating source \citep{trinidad2003}, is also connected to a disk (\citealt{vandertak2006, wang2012, gieser2019}) as well as large scale ($>1\arcmin$) outflow blue-shifted towards the south-west and red-shifted towards the north-east, traced in simple molecules (\citealt{Hasegawa1995, jimenezserra2012, vanderwiel2013, gieser2019}). \citet{vanderwiel2011} detect the methanol plume extending past this inner envelope towards the north-east. The CH$_3$OH 2$_{0,2,0}$-1$_{0,1,0}$ (E$_{\mathrm{up}}$\,=\,7~K) (Fig.~\ref{fig:afgl2591map}) line, which is blended with the 2$_{1,2,2}$-1$_{1,1,2}$ and 2$_{0,2,1}$-1$_{0,1,1}$ (E$_{\mathrm{up}}$\,=\,13, 20~K) lines, clearly traces this plume in multiple energy levels and at greater intensity than around VLA 1-5. The 6$_{0,6,0}$-5$_{0,5,0}$ (E$_{\mathrm{up}}$\,=\,48~K) line shows greatest intensity around VLA 1-5 but still extends into the NE, further supporting the findings from \citet{vanderwiel2011}. 

The colder temperatures, $<$20~K, for gas-phase CH$_3$OH indicate there are likely non-thermal desorption methods liberating this molecule after formation on dust grain surfaces. Cold methanol, at different scales and differing spatial features, is seen in numerous sources with differing mechanisms proposed. \citet{bouvier2020}, with single dish observations of Orion Molecular Cloud 2/3, conclude the detected CH$_3$OH is originating from the photo-dissociation regions rather than the hot corino.

\citet{vastel2014}, with single dish observations of the prestellar core L1544, conclude that non-thermal desorption methods, specifically photo-desorption, is responsible for the cold CH$_3$OH detected. \citet{Soma2015}, similarly in Taurus Molecular Cloud 1, offer several possibilities for non-thermal CH$_3$OH desorption: weak shocks, photoevaporation, and chemical desorption. \citet{favre2020} find a methanol `blob' at smaller scales (1000 AU) using interferometric observations of L1521F -- a very low luminosity object transitioning from prestellar to protostellar stages -- with the likeliest mechanism for desorption being a slow, gentle shock. In AFGL 2591, \citet{vanderwiel2011} propose this large-scale `plume' could be CH$_3$OH liberated by a shock front. The outflow (Fig.~14 of \citealt{vanderwiel2011}) coincides spatially with the production of this `plume.' In addition, it is seen in a range of E$_{\mathrm{up}}$ transition lines and thus is not discriminatory by temperature. Our component two aligns spatially with this feature. Our component three, also at T$_{ex}$\,=\,8~K, represents a blue-shifted structure to the north-east of VLA 3. \citet{gieser2019}, at small scales, suggest there is a young secondary outflow traced in SiO, and reason it is possible if there are numerous young stellar objects within VLA 3. \citet{suri2021}, subsequently, present NOEMA observations showing fragmentation of VLA 3 into three low-mass cores. In our observations, however, we do not have enough information to determine if CH$_3$OH component three is due to an associated outflow.

CH$_3$CCH has a range of E$_{\mathrm{up}}$ transition lines giving confidence in the LTE excitation temperature results. Similar to other star-forming regions, and as seen in WCCC sources, the carbon chains are on an extended scale with gas temperatures of around 30~K. In comparison, the COMs are not tracing the hot core at extended scales as such, and represent other mechanisms leading to their presence in the gas phase. Chemical modelling is needed to determine the formation routes, and is presented in Sect.~\ref{chemicalmodel}.

\begin{figure}[h!]
\centering
\includegraphics[trim=2cm 2cm 2cm 2cm, clip=true, width=0.5\textwidth]{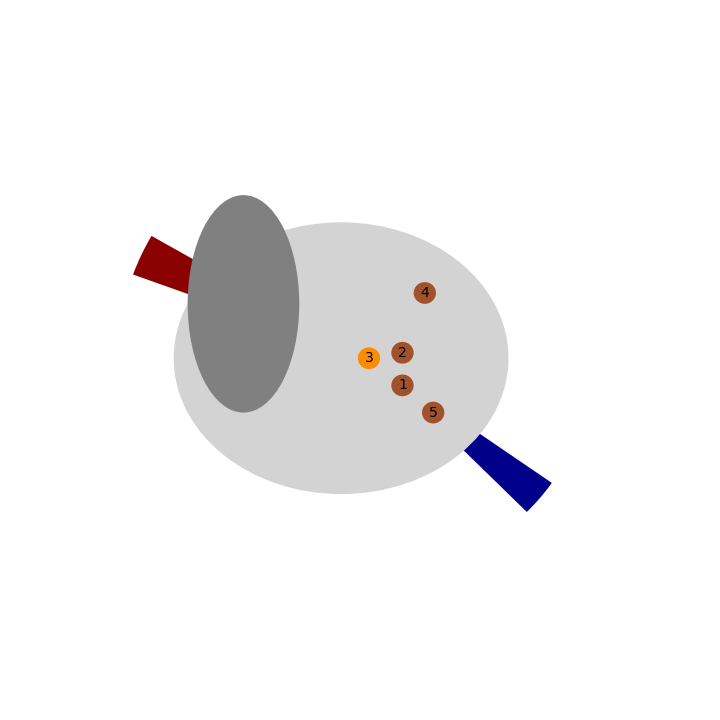}
\caption{Cartoon depiction of AFGL 2591. The brown circles represent the sources VLA 1-5, with the dominant protostellar source VLA 3 highlighted. The warm, dense inner envelope is in light grey, while the methanol plume is in dark grey. The general direction of the outflows are represented in red and blue, marking the shift in velocity as well.}
\label{fig:afgl2591cartoon}
\end{figure}

The comprehensive large scale structure of AFGL 2591, as suggested by our results and the literature, is shown graphically in Fig.~\ref{fig:afgl2591cartoon}. The brown circles represent the sources VLA 1-5, with the dominant and most massive hot core VLA 3 highlighted. The warm, dense inner envelope is in light grey, while the methanol plume is in dark grey. The general direction of the outflows are represented in red and blue, marking the shift in velocity as well.

\subsection{IRAS 20126}
\label{section:iras20126discussion}

IRAS 20126 is a relatively simple high-mass star-forming region. The peak of molecular emission, in both our integrated intensity (Fig.~\ref{fig:iras20126map}) and column density maps (Fig.~\ref{fig:iras20126paramresults}), is located near the protostellar source. The central molecular gas peaks at the site of the protostar and is oriented perpendicular to a small-scale outflow, involving a known disk (\citealt{cesaroni1997,Zhang1998,cesaroni1999,cesaroni2005}). The increased line widths (Fig.~\ref{fig:iras20126paramresults}) towards the central protostellar region can be hypothesised to be due to increased turbulence, as in \citet{cesaroni1997}. The found temperatures in this source are not high enough to produce thermal broadening -- it would be, for a methanol molecule at 30~K, only 0.4 km s $^{-1}$.

The protostar is young and embedded in a dense hot core \citep{cesaroni1997}, reaching up to $~$200~K in the core where COMs are known to exist (\citealt{cesaroni2005, Xu2012}. Our single dish observations are insufficient to spatially resolve the protostar, and we are seeing the protostar smoothed into the surrounding dense gas. \citet{cesaroni1997}, also with IRAM 30m observations, found a rotational temperature of 50~K for CH$_3$OH, similar to the peak in our maps of both CH$_3$OH and CH$_3$CCH.

In our observations, CH$_3$CCH and $c$-C$_3$H$_2$ have visibly different velocity orientations than CH$_3$OH and H$_2$CO; however, neither group are oriented with the dense molecular gas surrounding the disk but with each of the two known outflows. The disk orientation, at scales of about 1000 AU \citep{sridharan2005}, is also not seen in our beam of 16 000 AU. A large-scale (up to 2$\arcmin$) outflow traced in $^{12}$CO has a red-shifted northern flow and a blue-shifted southern flow (\citealt{wilking1990, shepherd2000}), corresponding to the same gradient and orientation seen in the $v_{lsr}$ of CH$_3$CCH with velocities increasing southward. The small-scale ($<$ 30$\arcsec$) bipolar outflow oriented north-west to south-east is traced in simple molecules (\citealt{cesaroni1997, cesaroni1999}), suggested to be a jet-driven bow shock \citep{Su2007}. 

In \citet{cesaroni1997} there are curious dynamics seen in the CS(3-2) and HCO+(1-0) transition lines with blue-shifted outer wings and red-shifted inner wings in the north-west, and red-shifted outer wings and blue-shifted inner wings in the south-east. They interpret the structure as due to the orientation of the outflow axis relative to the plane of the sky. This outflow is traced by CH$_3$OH and H$_2$CO in our observations with a blue-shifted south-east flow, indicating it arises similarly to the inner wings. \citet{palau2017} also find that COMs are seen extended towards the south-east in IRAS 20126, suggesting they arise from the warm cavity walls or post-shock region. \citet{palau2017}, in their modelling, find that H$_2$CO, while possible to form in the gas phase, is co-spatial with larger COMs and thus likely arises from the same mechanism. Chemical modelling for our results is presented in Sect.~\ref{chemicalmodel}.

\begin{figure}[h!]
\centering
\includegraphics[trim=2cm 2cm 2cm 2cm, clip=true, width=0.5\textwidth]{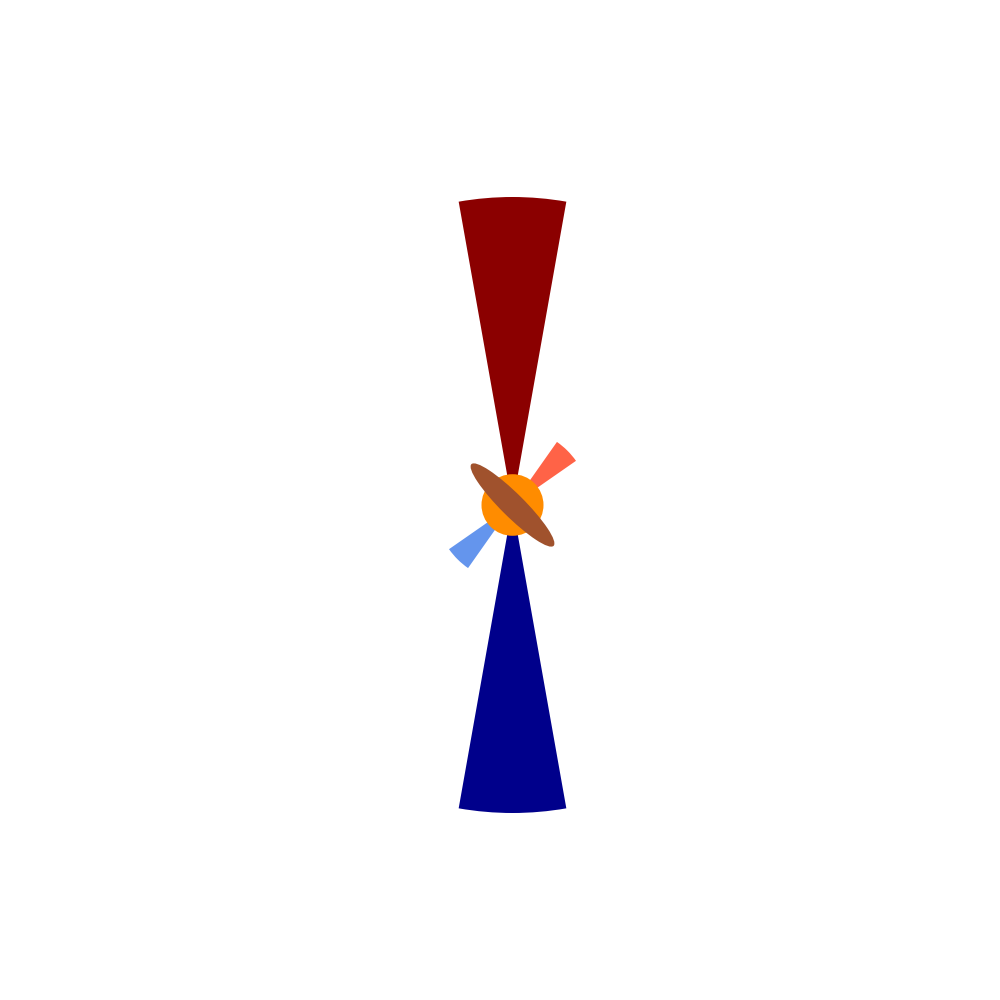}
\caption{Cartoon depiction of IRAS 20126. The orange circle represent the protostellar source, and the brown ellipse represents the central molecular gas and disk. The small-scale south-east--north-west outflow, traced by simple molecules, is in light red where red-shifted, and light blue where blue-shifted. The large-scale north-south outflow is similarly represented in dark red and dark blue.}
\label{fig:iras20126cartoon}
\end{figure}

As in AFGL 2591, at kpc distances our single dish observations are showing large-scale emission and not the hot core. While we cannot discern molecular presence at the hot core scale, and thus the differentiation between COMs and WCCC species, there is a notable difference in the physical conditions that the tracers CH$_3$OH and H$_2$CO arise in versus CH$_3$CCH and $c$-C$_3$H$_2$. These components, as seen in our observations and past studies, are summarised graphically in Fig.~\ref{fig:iras20126cartoon}.

In the above discussion for both IRAS 20126 and AFGL 2591, our results are limited by the number of lines detected compared to the number of physical components present in the sources. A one- or two-component fit is representative in certain cases but may not reveal composite structure. \citet{bouvier2020} are careful to note that a lot of complex molecule studies, which would include our own, are single dish observations and cannot spatially differentiate sub-structures of the regions.

These results would also have increased confidence with better S/N data, especially at high frequencies. There is a thorough E$_{\mathrm{up}}$ range, however, covered in the transition lines within our frequency ranges which lends confidence to the excitation temperatures found, especially when molecules are in LTE.

\subsection{Chemical model}
\label{chemicalmodel}

In order to provide concrete support for the chemical formation of these molecules in the observed environments, we modelled the species abundances of both high-mass star-forming regions using the NAUTILUS gas-grain code \citep{ruaud2016}. NAUTILUS simulates the chemical evolution of a region in three phases -- gas, grain surface, and grain mantle -- using coupled differential equations to describe a network of over 10,000 reactions connecting 489 species. With NAUTILUS, we model how the abundance of any given species changes over time as a function of different astrophysical conditions. These conditions can be static or dynamic throughout the simulation, and include properties such as gas density, dust and gas temperature, visual extinction, ultraviolet flux and cosmic ray ionisation.

In this discussion, we lead with and display the models for CH$_3$CCH and CH$_3$OH. These two species have more numerous strong detections relative to $c$-C$_3$H$_2$ and H$_2$CO and are representative of CCM and COM evolution, respectively.

\begin{table}
\caption{Initial ISM abundances in the NAUTILUS model.}
\label{tab:abundances}
\centering
\begin{tabular}{l c}
\hline\hline
Species & Abundance \\
\hline
He     &   9.0 $\times$ 10$^{-2}$ \\
N    &     6.2 $\times$ 10$^{-5}$ \\
O     &    2.4 $\times$ 10$^{-4}$ \\
H     &    0.0 \\
H2      &  0.5 \\
C+     &   1.7 $\times$ 10$^{-4}$ \\
S+     &   1.5 $\times$ 10$^{-5}$ \\
Si+    &   8.0 $\times$ 10$^{-9}$ \\
Fe+     &  3.0 $\times$ 10$^{-9}$ \\
Na+    &   2.0 $\times$ 10$^{-9}$ \\
Mg+     &  7.0 $\times$ 10$^{-9}$ \\
P+      &  2.0 $\times$ 10$^{-10}$ \\
Cl+     &  1.0 $\times$ 10$^{-9}$ \\
F      &   6.7 $\times$ 10$^{-9}$ \\
\hline
\end{tabular}
\end{table}

\subsubsection{AFGL 2591 chemical model results}

\subsubsection*{CH$_3$CCH and CH$_3$OH}

We first determined the fractional abundances of CH$_3$CCH and CH$_3$OH in the cloud -- the ratio between the column density of the analysed chemical to the column density of molecular hydrogen, H$_2$. We calculated the H$_2$ column density map for AFGL 2591 using archival JCMT SCUBA2 850~$\mu m$~emission (JCMTCAL observations on 2022/09/05). These maps were re-gridded in CASA\footnote{https://casa.nrao.edu/} to match the spatial resolution of the GBT and IRAM data. 

The 850~$\mu m$~continuum is caused by dust emission, which correlate to the H$_2$ column densities as molecular hydrogen generally forms on dust grains \citep{wakelam2017}. We used a gas to dust mass ratio of 100, a constant dust mass opacity coefficient of $\kappa$ = 1~cm$^2$/g appropriate for MRN dust with a thin ice mantle (MRN dust describes the dust size distribution in the Galaxy; \citealt{mathis1977, ossenkopf1994}), a mean mass per particle of $\mu = 2.32$, and a temperature derived from the LTE modelling of CH$_3$CCH and CH$_3$OH (typically $\approx$\,30~K). We divided the CH$_3$CCH LTE model column density map by the H$_2$ column density map and found CH$_3$CCH abundances varying from $5\times10^{-10}$ to $5\times10^{-9}$ with the higher abundances in the northernmost `blob' in Fig.~\ref{fig:afgl2591paramresults}.  

CH$_3$OH is more complicated with three distinct kinematic components. As we could not, a priori, determine how much of the total H$_2$ column density is associated with each component, we explored the CH$_3$OH abundances in three small regions where we could reasonably isolate each component -- a position in one component where there is no emission from the other two components. Since the LTE modelling of these three components utilised fixed temperatures of 18~K for component one and 8~K for components two and three, we used these temperatures to calculate the H$_2$ column densities in each of these components from the 850~$\mu m$~map. We divided the CH$_3$OH column density maps for each component by the appropriate H$_2$ column density map and found CH$_3$OH abundances of $6.3\times10^{-10}$ for component one (18~K), $3.7\times10^{-10}$ for component two (8~K), and $2.3\times10^{-10}$ for component three (8~K).

In NAUTILUS, we started with a two-phase model. First, a cold quiescent cloud was allowed to evolve for 10$^5$ years. This stage was run with initial abundances appropriate for the diffuse ISM (see Table~\ref{tab:abundances}), a temperature of 10~K, a density of $10^4$~cm$^{-3}$, and a visual extinction of 50 magnitudes. Then, we assume that embedded protostars warm up the cloud, a stage evolved for another 10$^6$ years. In the second part of this simulation, the observed temperature of each pixel found from the LTE models is used as the dust and gas temperatures (typically 20-30~K), while the visual extinction is calculated from the average H$_2$ column density. Studies of the H$_2$ to A$_V$ conversion range from A$_V$ = $\frac{N_{H_2}}{2.1\times10^{21}}$ (\citealt{rachford2009, zhu2017}) to A$_V$ = $\frac{N_{H_2}}{1.9\times10^{21}}$ (\citealt{bohlin1978, whittet1981}). In this paper, we assumed an `average' value of A$_V$ = $\frac{N_{H_2}}{2\times10^{21}}$. We used a standard cosmic ray ionisation rate of $1.3\times10^{-17}$ s$^{-1}$. For the UV field, the NAUTILUS code provided a function of the form $S\times10^8$~photons~cm$^{-2}$~s$^{-1}$ \citep{ruaud2016} where S is a scaling factor. We used a scaling factor of 1 to model a general interstellar radiation field. In following models, we modified both the CR ionisation rate and the UV scaling factor.

The volume densities of massive star-forming regions, on the scales that we present here, have been shown to have a centre-to-edge gradient of $n \propto r^{-\alpha}$ where $\alpha$ ranges from 1.0 to 1.6 (\citealt{vandertak2000,beuther2002}). To account for such density gradients, we calculated models for three different densities: 10$^4$ which is appropriate for the cloud edges, 5 $\times$ 10$^4$ an intermediate value, and 10$^5$~cm$^{-3}$ which simulates conditions in the cloud cores.

\begin{figure*}[h!]
\centering
\includegraphics[width=\textwidth]{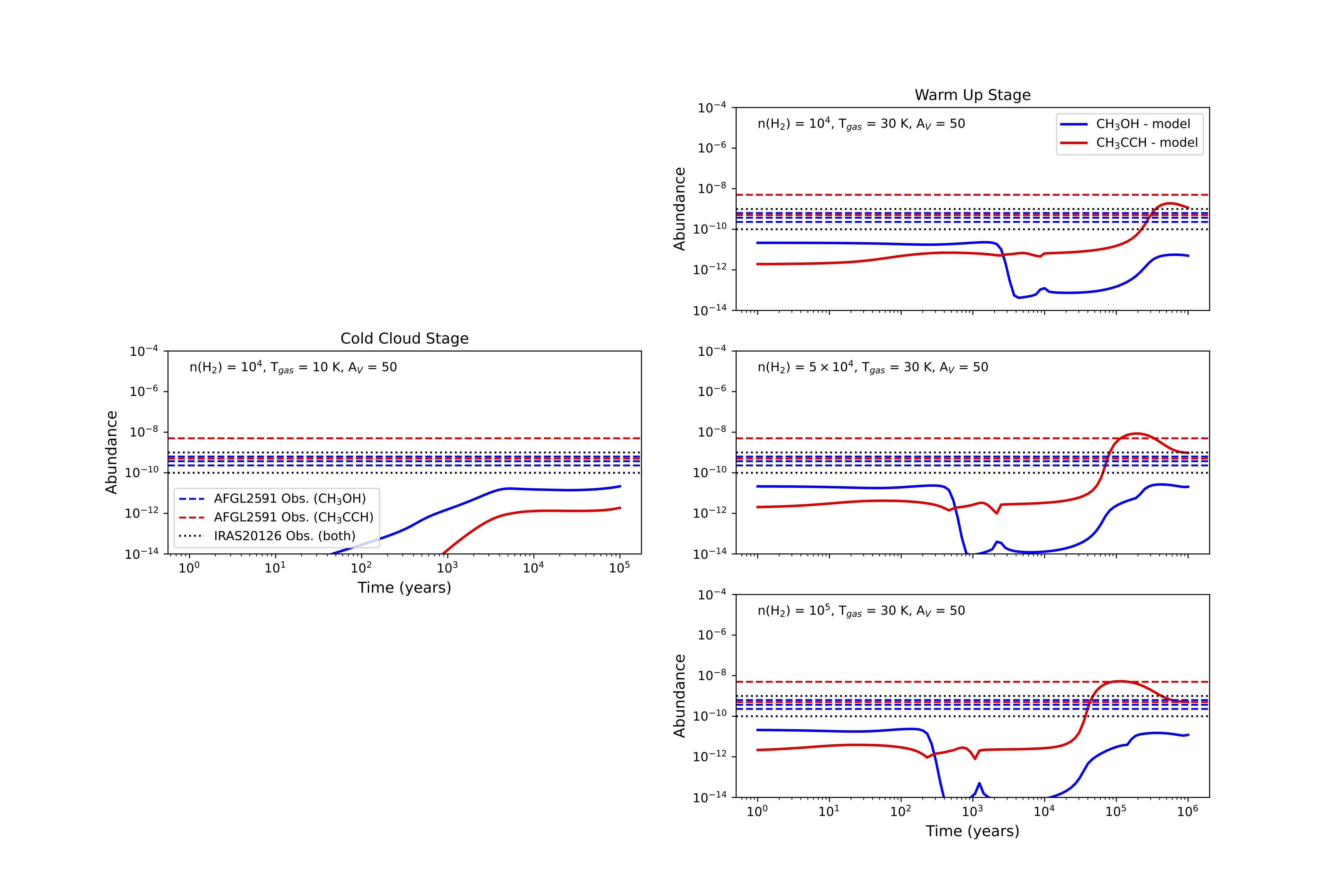}
\caption{Results from the NAUTILUS two stage chemical evolution model. The solid blue and red lines represent the time evolution of the CH$_3$OH and CH$_3$CCH abundances, respectively. The blue and red dashed horizontal lines indicate the minimum and maximum observed abundances for CH$_3$OH and CH$_3$CCH, respectively, in AFGL 2591. The dotted black line indicates the minimum and maximum observed abundances in IRAS 20126. The left panel shows the cold cloud evolution stage and the right panels show the warm-up stage for three different densities.}
\label{fig:afgl2stage}
\end{figure*}

Figure~\ref{fig:afgl2stage} shows the results of these connected simulations. The horizontal dashed blue and red lines show the minimum and maximum CH$_3$OH and CH$_3$CCH abundances, respectively. In general, this simple simulation can produce the observed CH$_3$CCH abundances for the majority of the cloud at about $10^6$ years. This is true regardless of position -- at the cloud edges where n = 10$^4$ or the cores where n = 10$^5$. The one caveat to this statement, however, occurs in the northern part of the map, where the CH$_3$CCH abundances are the highest. If the density in this region is as low as $10^4$~cm$^{-3}$, then our models cannot account for these high observed abundances. Figure~\ref{fig:afgl2stage} also shows that we cannot reproduce the observed CH$_3$OH abundances at anytime for any density.  

Given the poor match to our CH$_3$OH observations, we modified the chemical model to attempt to restrict the conversion of CH$_3$OH from the gas phase to the dust phase, or to help thermally desorb CH$_3$OH off the dust grains. A variety of methods were attempted, including increasing the ambient ultraviolet flux and cosmic ray ionisation rates. Increasing the scaling factor of the UV flux to S = 1000 had no effect upon the chemistry of either species. This is expected since the UV reaction rates take the form $k \propto e^{-S~A_V}$ and A$_V$ is generally so high that the UV field is completely attenuated. To simulate the effects of an external radiation field on the lower volume and column density cloud edges, we also ran a model with n = 10$^4$ and A$_V$ = 5. This simulation results in CH$_3$OH and CH$_3$CCH abundances that are respectively two and four orders of magnitude lower than those shown in Fig.~\ref{fig:afgl2stage} due to the increased rate of photodestruction of both molecules.

Increasing the cosmic ray ionisation rate by a factor of three increases the abundances of both species by a very small amount, but not enough to increase the CH$_3$OH abundances to observable values. Continued increases in the cosmic ray ionisation rate, however, have the opposite effect and result in lower abundances than those seen in Fig.~\ref{fig:afgl2stage}, again due to an increased destruction rate.

As thermal desorption mechanisms cannot produce the observed CH$_3$OH abundances, we turn our attention to non-thermal ones. A shockwave passing through a cloud filled with dust may induce sputtering, which causes the dust grains to shatter \citep{jones1994}, liberating the grain species back into the gas phase and causes a sharp increase in the gas phase abundances. An outflow, such as the ones seen in our sources, can produce such a shock (\citealt{zhang2005, herbst2009, palau2017}). Since it only includes thermal desorption processes, NAUTILUS cannot simulate sputtering directly. We simulated the effects of shock induced, non-thermal desorption (i.e. sputtering) by implementing a three stage model.

The first stage, cold cloud evolution, is the same as our prior model. The second stage, a shock, simulated the passage of a shock wave. We took the abundances at the end of stage one, removed all the grain surface species and added them back into the gas phase. The shock stage begun with enhanced gas phase abundances of all species and no grain surface species. During the shock we utilised the density and temperature evolution of the gas calculated by \citet{palau2017} for IRAS 20126 (see their Fig. 5, top).  This shock model follows the parametric approximation of \citet{jimenezserra2008} for C-type shocks, assuming a shock velocity of 40~km s$^{-1}$ and a pre-shock density of $10^4$~cm$^{-3}$. In our chemical models of this stage, we fixed the dust temperature to 80~K. The duration of the shock is $10^4$ years, during which the temperature sharply increased to 2500~K before decreasing to 50~K, while the density slowly increased from $10^4$ to $8.2\times10^4$~cm$^{-3}$, where it stabilised for the rest of the simulation. In the post shock stage, we returned the gas to the temperature used in the warm-up stage above, and allowed the cloud to evolve for a final $10^6$ years at a constant gas density of $8.2\times10^4$~cm$^{-3}$.

\begin{figure*}[h!]
\centering
\includegraphics[width=\textwidth]{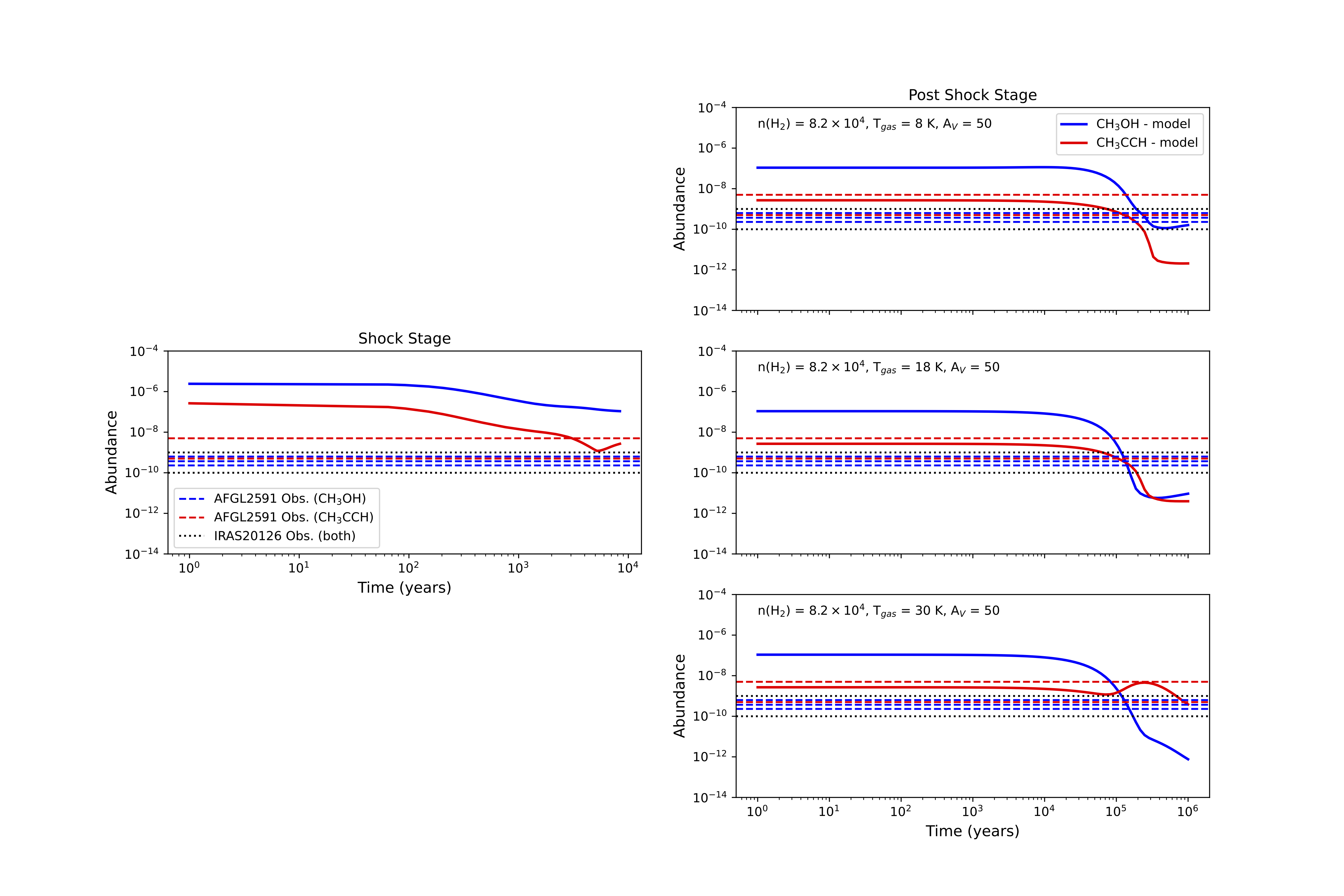}
\caption{Results from the NAUTILUS three stage chemical evolution model. The solid blue and red lines represent the time evolution of the CH$_3$OH and CH$_3$CCH abundances, respectively. The blue and red dashed horizontal lines indicate the minimum and maximum observed abundances for CH$_3$OH and CH$_3$CCH, respectively, in AFGL 2591. The dotted black line indicates the minimum and maximum observed abundances in IRAS 20126. The left panel shows the shock stage and the right panels show the post shock stage for three different temperatures.}
\label{fig:afgl3stage}
\end{figure*}

Figure~\ref{fig:afgl3stage} shows the results of the three stage chemical evolution model. The panel on the left shows the shock phase, and the panels on the right show the post shock evolution for temperatures of 8~K, appropriate for component two and component three of CH$_3$OH, 18~K, appropriate for component one of CH$_3$OH, and 30~K, appropriate for CH$_3$CCH.  As can be seen in the right panels, all of the observed abundances can be reached after about $10^5$~years of post shock evolution. For CH$_3$CCH, the higher abundance limit is reached much earlier in the simulation during the shock phase. In AFGL 2591, while the majority of the cloud can be modelled as a simple cold cloud followed by a warm up phase, as is appropriate for WCCC, the northern region may require a non-thermal desorption mechanism for CH$_3$CCH if the density is around $10^4$~cm$^{-3}$.

\subsubsection*{H$_2$CO and $c$-C$_3$H$_2$}

To model H$_2$CO and $c$-C$_3$H$_2$ we first assumed that the physical conditions were identical to those for CH$_3$CCH and CH$_3$OH; namely that the cloud experienced a 10~K cold stage followed by a 30~K warm-up stage.  We also used the H$_2$ column density map calculated at 30~K to determine the H$_2$CO and $c$-C$_3$H$_2$ abundances. Under these conditions, the H$_2$CO abundance ranges from $3\times10^{-9}$ in the cloud core to $10^{-8}$ in the north western cloud edges. The $c$-C$_3$H$_2$ abundances range from $4\times10^{-11}$ in the cloud core to $2\times10^{-10}$ along the western and northern edges. Under these conditions, the observed $c$-C$_3$H$_2$ abundances are reached in the cold cloud stage and remain high well into the warm-up stage. Note that the same is true if we model the $c$-C$_3$H$_2$ abundances using a temperature of 20~K to calculate the H$_2$ column density. H$_2$CO, however, requires the 30~K warm-up stage in order to reach the observed abundances.

While an `average' temperature of 30~K may be appropriate for $c$-C$_3$H$_2$ (see Fig.~\ref{fig:afgl2591paramresults}), it is not for H$_2$CO where the average temperature is about 40-50~K. For H$_2$CO we used temperatures of 50~K to calculate the H$_2$ column density and then the abundance map. Given the higher temperature, the resultant H$_2$ column densities are lower, resulting in slightly increased H$_2$CO abundances which now vary from  $5\times10^{-9}$  to $2\times10^{-8}$.  Under these conditions, the model cannot  produce the observed abundances in the cold cloud stage, but can easily do so early in the 50~K warm-up stage. Therefore, no shock liberation of either molecule off the grain surfaces is required to produced the observed abundances.

\subsubsection{IRAS 20126 chemical model results}

\subsubsection*{CH$_3$CCH and CH$_3$OH}

In IRAS 20126, we only considered the main component of the cloud since, other than the cold `foreground screen' described in Sect.~\ref{sec:iras20126ltemodel}, the CH$_3$CCH and CH$_3$OH LTE model is dominated by one main component. To produce the corresponding H$_2$ column density map, we used archival JCMT SCUBA-1 850~$\mu m$~continuum observations (Proposal ID M98BU24). These maps were re-gridded to match the spatial resolution of the GBT and IRAM data. We divided the CH$_3$CCH and CH$_3$OH LTE model column density maps by the H$_2$ column density map and found abundances varying from 10$^{-10}$ and 10$^{-9}$ for both species.

We followed the same chemical modelling approach in NAUTILUS as was used for AFGL 2591. We allowed the cloud to evolve for $10^5$ years under cold conditions (n$=10^4$~cm$^{-3}$, T = 10~K) starting with general ISM abundances, and then warmed the gas to 30~K to evolve for another $10^6$ years. 

Figure~\ref{fig:afgl2stage} shows the results of these simulations, along with the results for AFGL 2591. The left panel displays the cold cloud stage with the parameters listed above and the right panel shows the warm-up stage for a pixel with T = 30~K, A$_V$ = 50 (i.e. N(H$_2$) $= 10^{23}$~cm$^{-2}$), and, from the top, $n=10^4$~cm$^{-3}$, $n = 5\times10^4$~cm$^{-3}$, and $n = 10^5$~cm$^{-3}$. The dotted black horizontal lines indicate the observed abundance limits of $10^{-10}$ and $10^{-9}$. Figure~\ref{fig:afgl2stage} shows that although the observed CH$_3$CCH abundances can be reproduced after a few $\times10^5$ years in the the warm up phase, the model CH$_3$OH abundances never reach the observed values. While we show the results for one specific pixel, these results are generally true in each pixel across the entire source. The main reason the modelled CH$_3$OH abundances are so low is due to freeze out of gas phase methanol onto the dust grains. 

As discussed previously, a number of outflows and associated shocks have been identified and observed within IRAS 20126 (e.g. \citealt{shepherd2000, zhang2001}). Similarly to AFGL 2591, we simulated the effects of shock induced desorption with a three stage model. We included a shock stage in which all of the molecules frozen onto the dust (after the cold cloud evolution stage) are sputtered back into the gas phase, using the same gas density and temperature evolution. Then, after the shock stage, the cloud is allowed to continue to evolve at a density of $8.2 \times 10^4$~cm$^{-3}$ and at the currently observed temperatures derived from the LTE model.

Figure~\ref{fig:afgl3stage} presents the results from the shock on the left and post shock stages on the right, for our test pixel with a temperature of 30~K. The corresponding cold cloud stage is identical to Fig.~\ref{fig:afgl2stage}. The dashed horizontal lines indicate the observed abundance limits of $10^{-10}$ and $10^{-9}$. Throughout the duration of the shock, and for the first $10^5$ years of the post shock evolution, the CH$_3$OH abundance remains above the observed values.  After this, CH$_3$OH begins to refreeze onto the dust grain surfaces and, in doing so, passes through the range of observed abundances. While CH$_3$CCH does not require a shock to produce the observed abundances (see Fig.~\ref{fig:afgl2stage}), the passage of a shock can also produce the observed abundances after $10^6$ years.

Again, while the results presented are for the physical parameters in a single pixel, these results are generally true across the entire region. Shocks, which we suggest are produced by the observed outflows, are required to explain the observed CH$_3$OH abundances over the entirety of IRAS 20126. They are not, however, needed to produce the observed abundances of CH$_3$CCH. This is as seen in AFGL 2591, with the exception of the northern region. Our results for CH$_3$CCH in IRAS 20126 are uniformly consistent with Warm Carbon Chain Chemistry.

\subsubsection*{H$_2$CO and $c$-C$_3$H$_2$}

The modelling of H$_2$CO and $c$-C$_3$H$_2$ proceeded in a fashion similar to that done for AFGL 2591. We first assumed the same conditions as for CH$_3$OH and CH$_3$CCH, and then used the temperatures determined from our LTE model (see Fig.~\ref{fig:iras20126paramresults}). With the H$_2$ column density map calculated at 30~K we derived H$_2$CO abundance ranges from $6\times10^{-11}$ around the cloud edges to $3\times10^{-10}$ in the cloud core. The $c$-C$_3$H$_2$ abundance ranges from $3\times10^{-11}$ in the cloud core and eastern edges to $10^{-10}$ along the western edges. Under these conditions, the abundances of both molecules are reached in the cold cloud stage and remain high well into the warm-up stage.

From Fig.~\ref{fig:iras20126paramresults} we see that the temperature derived for $c$-C$_3$H$_2$ is about 8~K and fairly uniform across the map. For H$_2$CO, the temperature is about 80~K and also fairly uniform. We used these two temperatures to calculate the respective H$_2$ column density maps and found that, for T = 8~K, the $c$-C$_3$H$_2$ abundance varies from $4\times10^{-12}$ in the cloud core and eastern edges to $10^{-11}$ along the western edges. For T = 80~K, the H$_2$CO abundance ranges from $2\times10^{-10}$ around the cloud edges to $10^{-9}$ in the cloud core. The chemical models are able to produce these observed abundances as well. For $c$-C$_3$H$_2$ at 8~K, the models match the observations early in the cold cloud stage. For H$_2$CO at 80~K, the model matches the observations early in the 80~K warm-up stage -- although, the lowest abundances can be reached in the cold cloud stage as well. Therefore, as with AFGL 2591, no shock liberation of either molecule off the grain surfaces is required to produce the observed abundances.

\section{Summary}
\label{section:summary}

This paper covers a targeted survey of carbon chain molecules in two high-mass star-forming regions -- AFGL 2591 and IRAS 20126 -- towards the Cygnus X star-forming complex. With observations from the IRAM 30 m telescope and the GBT 100 m dish, we present:

\begin{enumerate}
  \item detections of numerous rotational transition lines of two carbon chain molecules, CH$_3$CCH and $c$-C$_3$H$_2$, and two molecules related to the complex organic family, CH$_3$OH and H$_2$CO, in both sources.
  \item a LTE model, developed in python to loop over several spectra in a map, that determines the physical environment the detected molecular species arise from.
  \item findings of carbon chain and complex organic molecules originating from different gases. The excitation temperature of carbon chains is typically 20-30~K while for complex organics can range from 8-85~K at these large scales (smoothing out the hot core). The velocity structures of each molecular type vary, and trace known structures in the regions.
    \item chemical evolution modelling with NAUTILUS that demonstrates that the observed abundances of CH$_3$CCH can be produced in the warm-up stage, where the embedded protostar has warmed the cloud to temperatures of 20-30~K. This is consistent with Warm Carbon Chain Chemistry. CH$_3$OH, on the other hand, requires a non-thermal desorption method to produce the observed abundances. A shock, sputtering the molecules off the dust grain surface, adequately does this. These mechanisms are found for both high-mass star-forming regions.
\end{enumerate}

These results expand on current knowledge of carbon chain chemistry in high-mass star-forming regions. Warm carbon chain chemistry, and other envelope scale chemical mechanisms, can readily be studied with single dish telescopes. For an investigation of hot core chemistry, higher spatial resolution observations are required. These results also demonstrate the use of LTE models for mapping the physical environment using wide-band data cubes -- the type of data many modern telescopes are capable of providing.

\newpage

\begin{acknowledgements}

This work is based on observations carried out under project numbers 021-20 and 122-20 with the 30m telescope. IRAM is supported by INSU/CNRS (France), MPG (Germany) and IGN (Spain). The Green Bank Observatory is a facility of the National Science Foundation operated under cooperative agreement by Associated Universities, Inc.

PF would like to thank Larry Morgan and Dave Frayer at the GBT/NRAO for their help developing observing scripts as well as GBTIDL scripts for ARGUS calibration and reduction. PF would also like to thank Pablo Torne and Monica Rodriguez at IRAM for their help observing with the IRAM telescope.

PF and RP acknowledge the support of the Natural Sciences and Engineering Research Council of Canada (NSERC), through the Canada Graduate Scholarships – Doctoral, Michael Smith Foreign Study Supplement, and Discovery Grant programs.

As researchers at the University of Calgary, PF and RP acknowledge and pay tribute to the traditional territories of the peoples of Treaty 7, which include the Blackfoot Confederacy (comprised of the Siksika, the Piikani, and the Kainai First Nations), the Tsuut’ina First Nation, and the Stoney Nakoda (including Chiniki, Bearspaw, and Goodstoney First Nations). The City of Calgary is also home to the Métis Nation of Alberta Region 3.

\end{acknowledgements}

\bibliographystyle{aa} 
\bibliography{bib}

\begin{appendix}
\onecolumn
\section{Additional LTE model description}

\begin{table*}[h!]
\caption{Parameter inputs for the LTE model.}
\label{tab:ltemodelinput}
\centering
\begin{tabular}{ l c c c c }
\hline\hline
Species & N$_{tot}$ & T$_{ex}$ & v$_{lsr}$ & FWHM \\
 & (cm$^{-2}$) & (K) & (km s$^{-1}$) & (km s$^{-1}$) \\
\hline
AFGL 2591 & & & & \\
CH$_3$CCH& 1 $\times$ 10$^{11}$, 1 $\times$ 10$^{17}$, 1 $\times$ 10$^{14}$ & 3, 100, 40 & -10, -2, -5.7 & 0.5, 4, 2 \\
$c$-C$_3$H$_2$& 1 $\times$ 10$^{11}$, 1 $\times$ 10$^{17}$, 1 $\times$ 10$^{14}$ & 3, 100, 40 & -10, -2, -5.7 & CH$_3$CCH FWHM \\
H$_2$CO & 1 $\times$ 10$^{11}$, 1 $\times$ 10$^{17}$, 1 $\times$ 10$^{14}$ & 3, 150, 90 & -10, -2, -6 & 1, 7, 4 \\
CH$_3$OH C1 & 1 $\times$ 10$^{11}$, 1 $\times$ 10$^{17}$, 1 $\times$ 10$^{14}$ & 18 & -10, -1, -6 & 1, 7, 3 \\
CH$_3$OH C2 &  1 $\times$ 10$^{11}$, 1 $\times$ 10$^{17}$, 1 $\times$ 10$^{14}$ & 8 & -10, -2, -6 & 3.9 \\
CH$_3$OH C3 & 1 $\times$ 10$^{11}$, 1 $\times$ 10$^{17}$, 1 $\times$ 10$^{14}$ & 8 & -9.4 & 2.5 \\
\hline
IRAS 20126 & \\
CH$_3$CCH & 1 $\times$ 10$^{11}$, 1 $\times$ 10$^{17}$, 1 $\times$ 10$^{14}$ & 3, 100, 30 & -5, -1, -3.1 & 0.5, 4, 2 \\
$c$-C$_3$H$_2$ C1 & 1 $\times$ 10$^{11}$, 1 $\times$ 10$^{17}$, 1 $\times$ 10$^{14}$ & 3, 100, 10 & -5, -1, -3.1 & 0.5, 4, 2.4 \\
 $c$-C$_3$H$_2$ C2 & 1.0 $\times$ 10$^{13}$ & 3, 100, 10 & C1 v$_{lsr}$ & 0.5 \\
H$_2$CO     & 1 $\times$ 10$^{11}$, 1 $\times$ 10$^{17}$, 1 $\times$ 10$^{14}$ & 3, 150, 50 & -6, -1, -3.6 & 1, 8, 6.9 \\
CH$_3$OH C1 & 1 $\times$ 10$^{11}$, 1 $\times$ 10$^{17}$, 1 $\times$ 10$^{14}$ & 3, 150, 50 & -6, -1, -3.6 & 1, 8, 6.9 \\
CH$_3$OH C2 &  2.1 $\times$ 10$^{13}$ & 10 & C1 v$_{lsr}$-0.2 & 1.2 \\
\hline
\end{tabular}
\tablefoot{Where the parameter is variable, the inputs are minimum, maximum, and starting value, and where the parameter is fixed, the input is the starting value. If a parameter is linked to another, the reference parameter is noted in the table. The final parameter, source size, is always fixed at 20$\arcsec$.}
\end{table*}

\begin{figure*}[h!]
     \centering
     \begin{subfigure}[b]{0.57\textwidth}
         \centering
         \includegraphics[width=\textwidth]{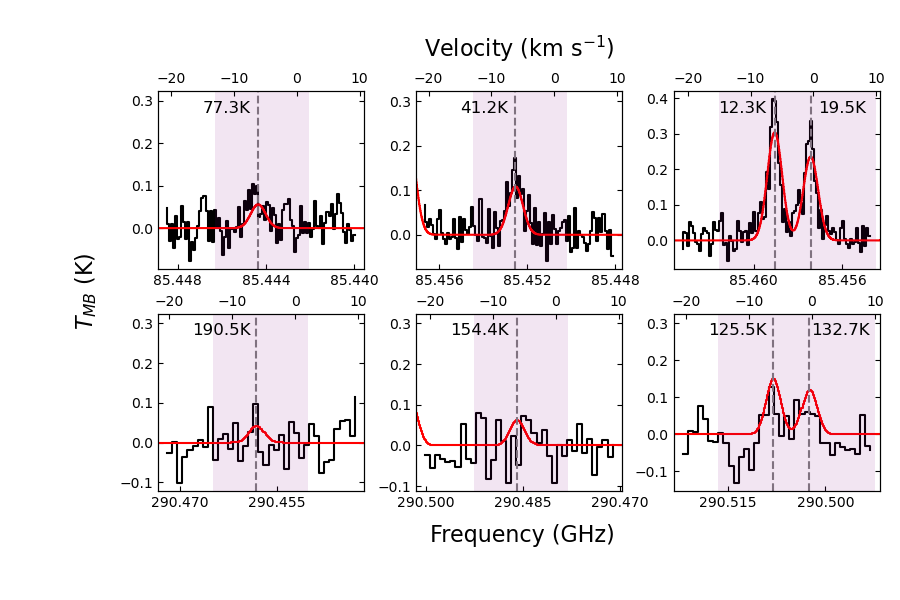}
         \caption{}
     \end{subfigure}
     \hfill
     \begin{subfigure}[b]{0.41\textwidth}
         \centering
         \includegraphics[width=\textwidth]{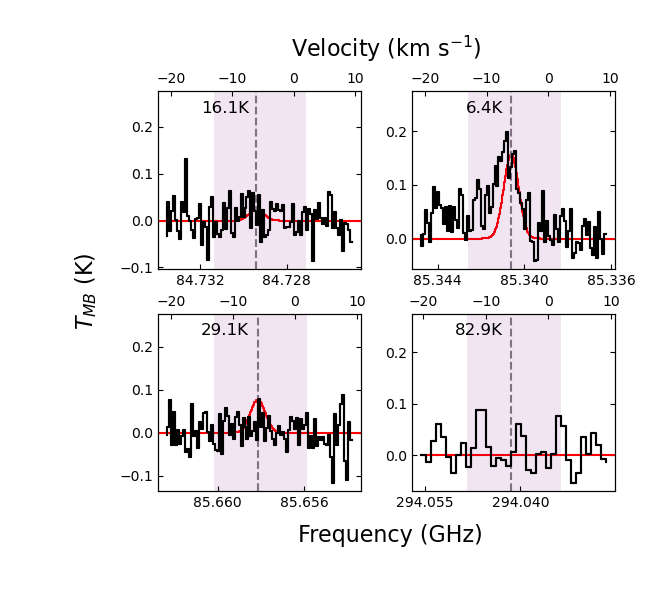}
         \caption{}
     \end{subfigure}
        \caption{Spectral line fits from the LTE model in AFGL 2591. All transition lines are shown of (a) CH$_3$CCH on the brightest pixel of the CH$_3$CCH 5$_0$-4$_0$ (E$_{\mathrm{up}}$\,=\,12~K), line, and (b) $c$-C$_3$H$_2$ on the brightest pixel of the $c$-C$_3$H$_2$ 2$_{1,2}$-1$_{0,1}$ (E$_{\mathrm{up}}$\,=\,6~K), line. The model spectrum is in colour while the observed data is in black. One component is used to fit the two carbon chain molecules together. The dashed grey line represents the observed transition frequency, with the energy level of this transition labelled.}
        \label{fig:afgl2591ccmresults}
\end{figure*}

\begin{figure*}
     \centering
     \begin{subfigure}[b]{0.735\textwidth}
         \centering
         \includegraphics[width=\textwidth]{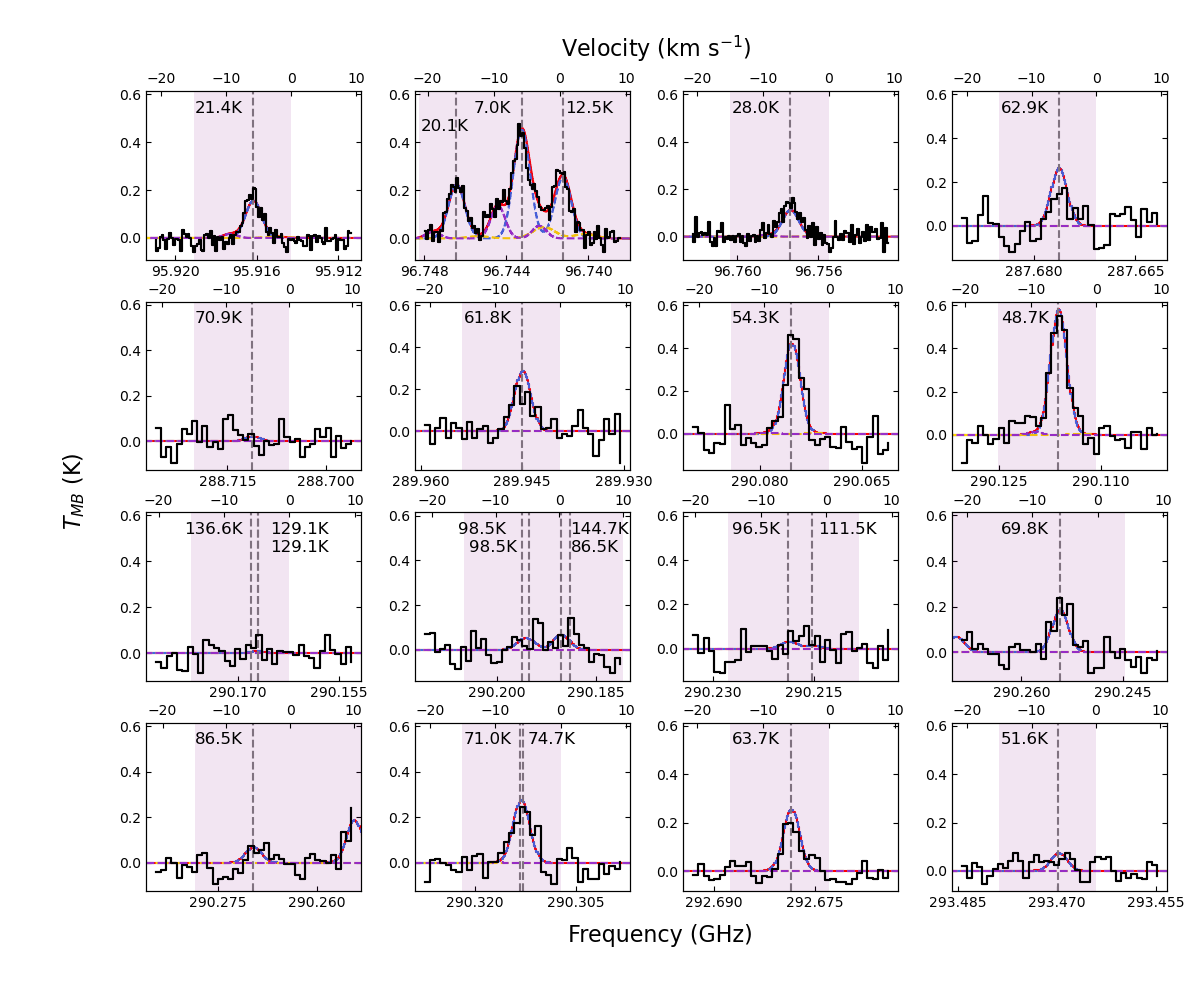}
         \caption{}
     \end{subfigure}
     \hfill
     \begin{subfigure}[b]{0.245\textwidth}
         \centering
         \includegraphics[width=\textwidth]{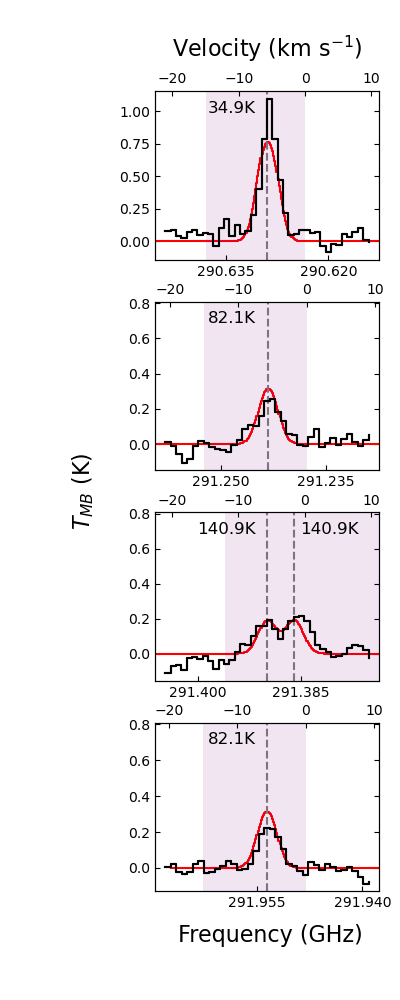}
         \caption{}
     \end{subfigure}
        \caption{Spectral line fits from the LTE model in AFGL 2591. All transition lines are shown of (a) CH$_3$OH on the brightest pixel of the CH$_3$OH on the brightest pixel of the CH$_3$OH 6$_{0,6,0}$-5$_{0,5,0}$ (E$_{\mathrm{up}}$\,=\,48~K) line, and (b) H$_2$CO on the brightest pixel of the H$_2$CO 4$_{0,4}$-3$_{0,3}$ (E$_{\mathrm{up}}$\,=\,35~K) line. The model spectrum is in colour while the observed data is in black. Three components are used to fit CH$_3$OH while one component is used to fit H$_2$CO. The dashed grey line represents the observed transition frequency, with the energy level of this transition labelled.}
        \label{fig:afgl2591comresults}
\end{figure*}

\begin{figure*}[h!]
\centering
\includegraphics[width=0.5\textwidth]{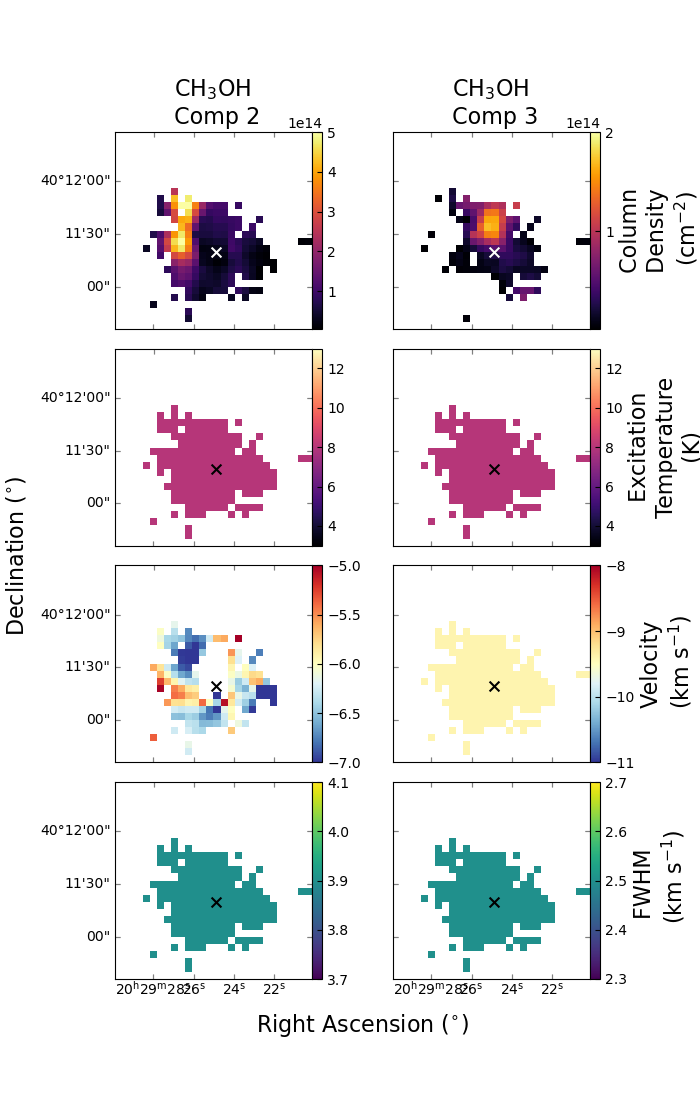}
\caption{Parameters maps for components two and three of CH$_3$OH in AFGL 2591 from the LTE model. The black or white `x' in each plot represents the source VLA 3. For both components, T$_{ex}$ is fixed at 8~K. For component two, the FWHM is fixed at 3.9~km s$^{-1}$. For component three, the v$_{lsr}$ is fixed at -9.4~km s$^{-1}$ and the FWHM at 2.5~km s$^{-1}$}
\label{fig:afgl2591paramresults2}
\end{figure*}

\begin{figure*}
     \centering
     \begin{subfigure}[b]{0.57\textwidth}
         \centering
         \includegraphics[width=\textwidth]{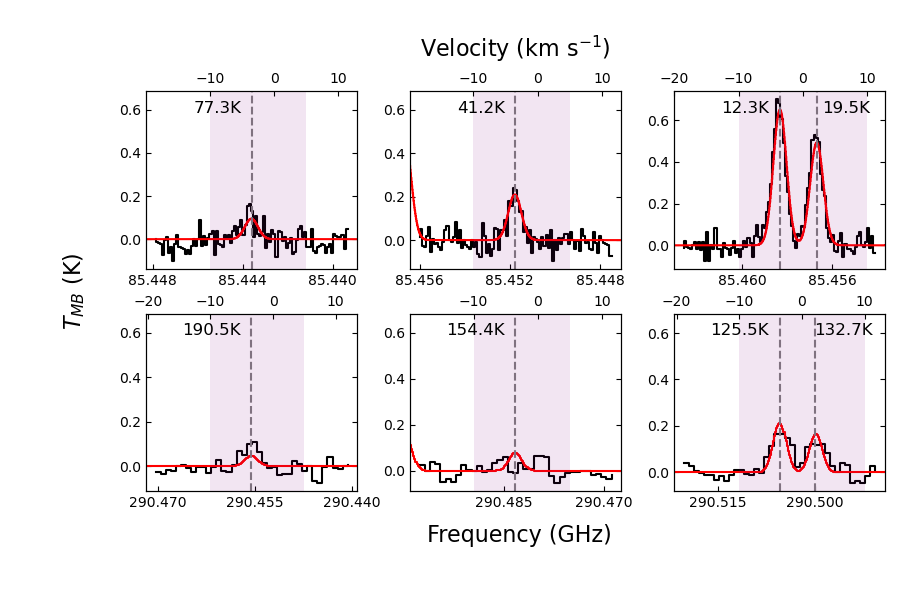}
         \caption{}
     \end{subfigure}
     \hfill
     \begin{subfigure}[b]{0.41\textwidth}
         \centering
         \includegraphics[width=\textwidth]{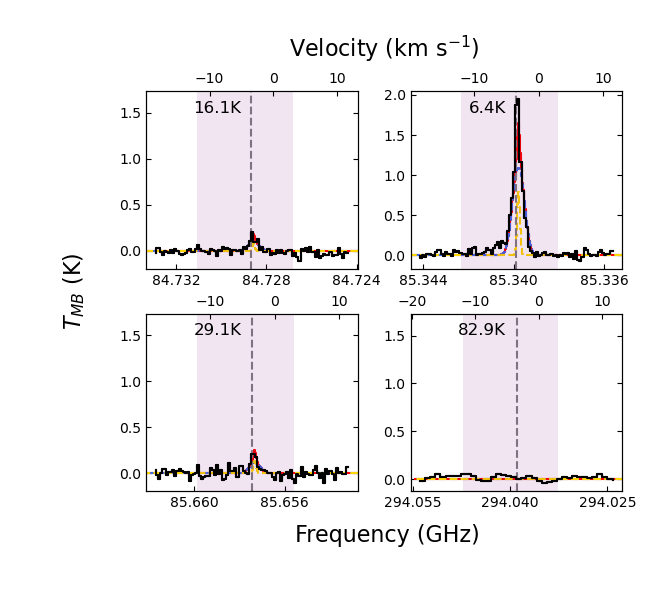}
         \caption{}
     \end{subfigure}
        \caption{Same as Fig.~\ref{fig:afgl2591ccmresults} for (a) CH$_3$CCH and (b) $c$-C$_3$H$_2$ in IRAS 20126. One component is used to fit CH$_3$CCH while two components are used for $c$-C$_3$H$_2$.}
        \label{fig:iras20126ccmresults}
\end{figure*}

\begin{figure*}
     \centering
     \begin{subfigure}[b]{0.735\textwidth}
         \centering
         \includegraphics[width=\textwidth]{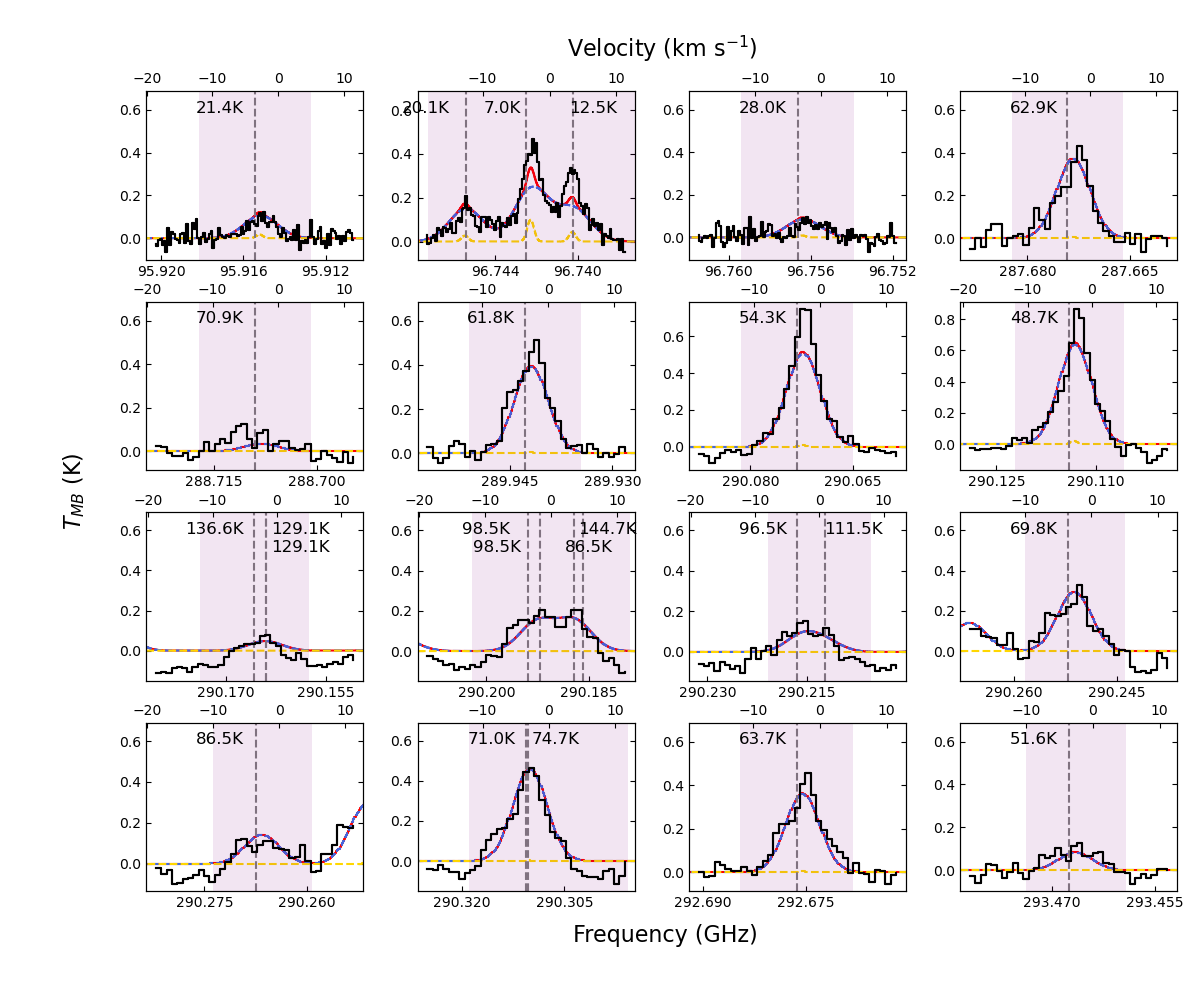}
         \caption{}
     \end{subfigure}
     \hfill
     \begin{subfigure}[b]{0.245\textwidth}
         \centering
         \includegraphics[width=\textwidth]{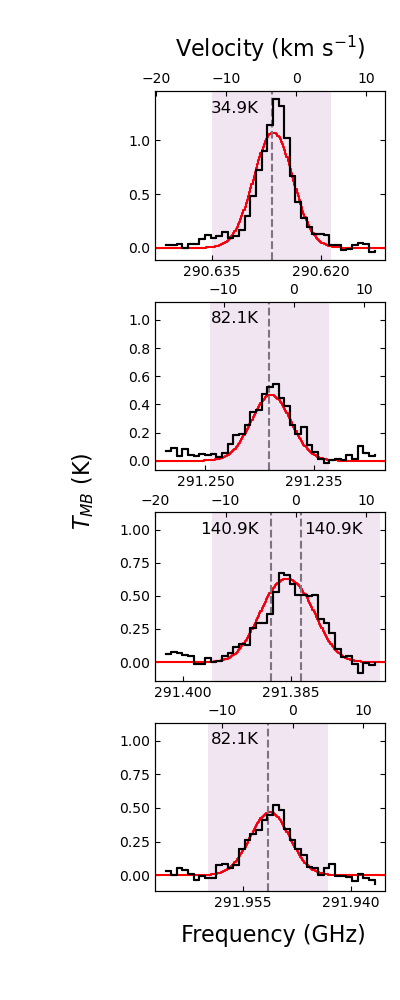}
         \caption{}
     \end{subfigure}
        \caption{Same as Fig.~\ref{fig:afgl2591comresults} for (a) CH$_3$OH and (b) H$_2$CO in IRAS 20126. Two components are used to fit CH$_3$OH while one component is used to fit H$_2$CO.}
        \label{fig:iras20126comresults}
\end{figure*}

\begin{figure*}[h!]
\centering
\includegraphics[width=0.5\textwidth]{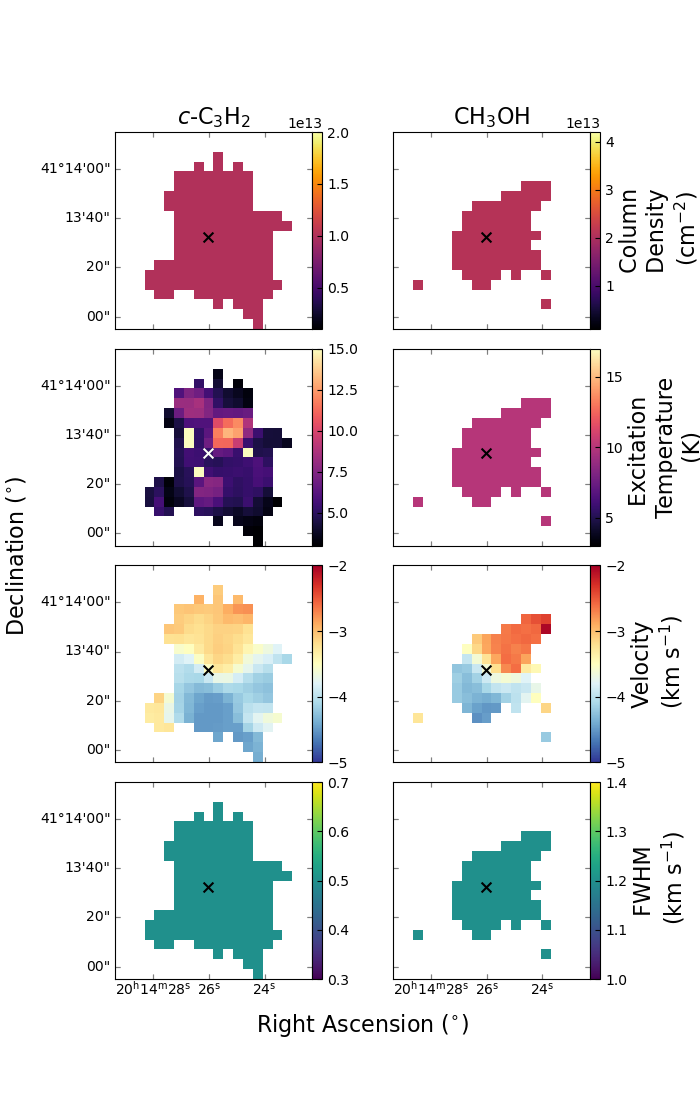}
\caption{Parameters maps for component two of $c$-C$_3$H$_2$ and CH$_3$OH in IRAS 20126 from the LTE model. The black or white `x' in each plot represents the protostar. For $c$-C$_3$H$_2$, N$_{tot}$ is fixed at 1 $\times$ 10$^{13}$~cm$^{-2}$ and the FWHM at 0.5 km s$^{-1}$. For CH$_3$OH, N$_{tot}$ is fixed at 2.1 $\times$ 10$^{13}$~cm$^{-2}$, T$_{ex}$ at 10~K the FWHM at 1.2 km s$^{-1}$.}
\label{fig:iras20126paramresults2}
\end{figure*}

\end{appendix}

\end{document}